\PassOptionsToPackage{table,dvipsnames}{xcolor}
\documentclass[11pt]{article}
\usepackage{amsfonts,slashed}
\usepackage{amsmath,amssymb,mathtools,mathrsfs,bm,booktabs,array,longtable,microtype}
\usepackage{arydshln}
\usepackage[greek,english]{babel}
\usepackage{cancel}
\usepackage{upgreek}
\usepackage{float}
\usepackage{url}
\usepackage{latexsym}
\usepackage{epsfig}
\usepackage{latexsym,amssymb}
\usepackage{amsmath,amssymb,amsthm,mathtools}
\usepackage{hyperref}
\usepackage{fontenc}
\usepackage{graphicx,tikz}
\usepackage{multirow}
\usepackage{xfrac}
\usepackage[T1]{fontenc}
\usepackage{textcomp}
\usepackage{lmodern}
\usepackage{stmaryrd}
\usepackage{cite}
\usepackage[normalem]{ulem}
\usepackage{placeins}   

\usepackage[vcentermath]{youngtab}

\newdimen\squaresize \squaresize=12pt
\newdimen\thickness \thickness=0.7pt

\def\square#1{\hbox{\vrule width \thickness
   \vbox to \squaresize{\hrule height \thickness\vss
      \hbox to \squaresize{\hss#1\hss}
   \vss\hrule height\thickness}
\unskip\vrule width \thickness} \kern-\thickness}

\def\cut#1{\hbox{\vrule width-1 \thickness
   \vbox to \squaresize{\hrule height-1 \thickness\vss
      \hbox to \squaresize{\hss#1\hss}
   \vss\hrule height-1\thickness}
\unskip\vrule width +4 \thickness} \kern-\thickness}

\def\vsquare#1{\vbox{\square{$#1$}}\kern-\thickness}

\usepackage{stmaryrd}

\usepackage[margin=20pt,small]{caption}
\usepackage{subcaption}

\usepackage{ifpdf}
\ifx\pdfoutput\undefined
   \pdffalse
   \usepackage{cite}
 \else
   \pdfoutput=1
   \pdftrue
  \ifdefined\pdfcompresslevel\pdfcompresslevel=9\fi
\fi
\DeclareFontFamily{OMS}{rsfs}{\skewchar\font'60}
\DeclareFontShape{OMS}{rsfs}{m}{n}{<-5>rsfs5 <5-7>rsfs7 <7->rsfs10 }{}
\DeclareSymbolFont{rsfs}{OMS}{rsfs}{m}{n}
\DeclareSymbolFontAlphabet{\Scr}{rsfs}

\numberwithin{equation}{section}
\def\be{\begin{equation}}
\def\ee{\end{equation}}
\def\ba{\begin{array}}
\def\ea{\end{array}}

\newcommand{\bea}{\begin{eqnarray}}
\newcommand{\eea}{\end{eqnarray}}

\def\={~=~}
\def\*{{}^*}

\def\eps{\epsilon}

\def\Tr{\mathrm{Tr}}

\usepackage[shadow,textwidth=2.7cm]{todonotes}
\usepackage{ifthen}
\reversemarginpar
\newcounter{todocounter}

\colorlet{fccolor}{blue!40!white}

\newcommand{\fcinline}[2][]{
  \ifthenelse { \equal {#1} {} }
    { \def\temp {#2} }  
    { \def\temp {#1} }   
  \refstepcounter{todocounter}\todo[color=fccolor,inline,caption={\textbf{\thetodocounter. FC} \temp}]{\textbf{\thetodocounter. FC:} #2}{}}

\colorlet{hscolor}{orange!20!white}

\newcommand{\hsinline}[2][]{
  \ifthenelse { \equal {#1} {} }
    { \def\temp {#2} }  
    { \def\temp {#1} }   
  \refstepcounter{todocounter}\todo[color=hscolor,inline,caption={\textbf{\thetodocounter. HS} \temp}]{\textbf{\thetodocounter. HS:} #2}{}}

\usepackage[dvipsnames]{xcolor}
\usepackage{tikz}
\usetikzlibrary{arrows.meta,positioning,calc}
\newcommand{\rep}[2]{(\mathbf{#1},\mathbf{#2})}
\newcommand{\Om}{\Omega}
\tikzset{
 box/.style={draw,rounded corners,minimum width=23mm,minimum height=8mm,fill=blue!4},
 nullbox/.style={box,dashed,draw=gray,fill=gray!8,text=gray!70!black},
 fbox/.style={box,fill=orange!10},
 lead/.style={-{Stealth[length=2.7mm,width=2mm]},thick,MidnightBlue},
 back/.style={-{Stealth[length=2.7mm,width=2mm]},thick,dashed,Orange},
 null/.style={gray,dotted,thick,-{Stealth[length=2.7mm,width=2mm]}},
 title/.style={font=\large\bfseries}
}
\newcommand{\pairarrows}[4][]{%
 \draw[lead,bend left=#4,#1] (#2) to (#3);
 \draw[back,bend left=#4,#1] (#3) to node[pos=.48,fill=white,inner sep=1pt]{\scriptsize$\Om$} (#2);}

\def\={~=~}
\def\*{{}^*}

\def\eps{\epsilon}

\def\Tr{\mathrm{Tr}}

\newcommand{\cF}{{\cal F}}

\newcommand{\cT}{{\cal T}}

\newcommand{\Gpolar}{\Gamma_{\!\#}}

\newcommand{\initX}{\mathsf{x}}
\newcommand{\initY}{\mathsf{y}}

\newcommand{\indI}[1]{%
  \ifcase#1 ?%
  \or {\textsc{i}}%
  \or {\textsc{j}}%
  \or {\textsc{k}}%
  \or {\textsc{l}}%
  \or {\textsc{m}}%
  \else ?%
  \fi
}
\newcommand{\indA}[1]{%
  \ifcase#1 ?%
  \or {\textsc{a}}%
  \or {\textsc{b}}%
  \or {\textsc{c}}%
  \or {\textsc{d}}%
  \or {\textsc{e}}%
  \or {\textsc{f}}%
  \else ?%
  \fi
}

\begin{document}
\allowdisplaybreaks[1]
\begin{titlepage}
\vfill
\begin{flushright}
\end{flushright}

\bigskip

\begin{center}
	{\LARGE \bf Polarized IKKT holography: \\[1ex]bulk realization of BPS operators}\\[1cm]

	{\large\bf Franz Ciceri\,$^{a}{\!}$
		\footnote{\tt fciceri@lpthe.jussieu.fr},
		 Aymeric Proust\,${}^{b}{\!}$
		\footnote{\tt aymeric.proust@ens-lyon.fr}  Henning Samtleben\,${}^{b,c}{\!}$
		\footnote{\tt henning.samtleben@ens-lyon.fr} \vskip .8cm}	
		{\it ${}^a$ Okinawa Institute of Science and Technology,\\1919-1 Tancha, Onna-son, Okinawa 904-0495, Japan}\\[2ex] 
		{\it ${}^b$ ENSL, CNRS, Laboratoire de physique, F-69342 Lyon, France}\\[2ex] \ 
	{\it  $^{c}$ Institut Universitaire de France (IUF)}\\ 
	
\end{center}
\vspace{0.5cm}

\noindent

\begin{abstract}
\noindent 
We study the holographic imprint of polarization in the lowest BPS multiplet of the IKKT matrix model. On the matrix model side, the deformation requires operator mixing and broken-symmetry Ward identities to organize the surviving operators into an irreducible atypical multiplet of the $F(4)$ superalgebra. In the bulk, we use one-dimensional maximal gauged supergravity to identify the analytic half-BPS background with $F(4)$ symmetry whose boundary behavior matches the IKKT polarization. We solve the linearized bosonic and fermionic equations around this background and its undeformed counterpart exactly, and extract the independent asymptotic data. The supersymmetry transformations of these data reproduce the matrix model multiplets, thus establishing a bulk realization of this BPS operator structure. 
\end{abstract}

\setcounter{footnote}{0}

\end{titlepage}


\tableofcontents \noindent {}

\section {Introduction}\label{sec:intro}


The IKKT matrix model was originally proposed as a nonperturbative formulation of type IIB string theory \cite{Ishibashi:1996xs}. Its elementary variables are matrices rather than fields on a prescribed spacetime, and the emergence of geometry is part of the dynamics.
This simplicity also makes the model difficult to analyze. The bosonic action is quartic and has flat directions, so expansion about a preferred configuration is not available a priori. After rescaling the matrices, the undeformed integral has no adjustable coupling with which to organize a semiclassical expansion.

Early work included exact calculations of the matrix integral \cite{Suyama:1997ig,Green:1997tn,Moore:1998et} and numerical studies of its partition function and eigenvalue distributions \cite{Krauth:1998xh,Krauth:1998yu,Krauth:1999qw}. The flat directions give rise to power-law tails in these distributions \cite{Krauth:1999qw}, raising separate convergence questions for the partition function and its observables \cite{Austing:2001pk}. The emergence of spacetime and possible spontaneous breaking of the SO(10) symmetry have been studied through effective eigenvalue dynamics, numerical simulations, and Gaussian expansions, with the complex fermion Pfaffian posing a persistent challenge \cite{Aoki:1998vn,Ambjorn:2000dx,Nishimura:2000ds,Nishimura:2001sx,Nishimura:2011xy,Anagnostopoulos:2022dak,Kumar:2022giw}.

An interesting path into these questions comes from the supersymmetric deformation of the model introduced in \cite{Bonelli:2002mb}, known as the polarized IKKT model. The deformation breaks the global ${\rm SO}(10)$ symmetry down to ${\rm SO}(3)\times{\rm SO}(7)$, preserving all supersymmetries. It plays a role analogous to the BMN mass deformation of BFSS matrix quantum mechanics \cite{Berenstein:2002jq} and to the ${\cal N}=1^*$ mass deformation of four-dimensional ${\cal N}=4$ super Yang-Mills theory \cite{Polchinski:2000uf}. Besides a cubic Myers term, the deformation adds quadratic terms for the bosonic matrices and a fermion mass term. The quadratic terms are a crucial difference from the undeformed IKKT model, which has only cubic and quartic interactions. They allow for the definition of a Gaussian approximation and propagators around the trivial configuration. The bosonic mass terms also lift the commuting flat directions of the undeformed action. The deformation parameter $\Omega$ controls the relative strength of the quadratic and interaction terms, providing a tunable handle on the integral. 
Polarization also changes the symmetry algebra. In the undeformed IKKT model the supercharges close into matrix-gauge transformations and hence anticommute to zero on gauge-invariant operators. In the polarized model they close into the bosonic ${\rm SO}(3)\times{\rm SO}(7)$ rotations of the simple Lie superalgebra $F(4)$. This nontrivial algebra provides a far more restrictive framework for organizing protected BPS operators and their multiplets. These features have enabled localization calculations, statistical analyses, and Monte Carlo studies of the deformed matrix integral \cite{Hartnoll:2024csr,Komatsu:2024ydh,Hartnoll:2025ecj,Chou:2025rwy}.

The polarized model has also been studied in the context of holography. A Euclidean type IIB description relates the fuzzy-sphere saddle to polarized branes \cite{Hartnoll:2024csr}, while backreacted supergravity geometries have been matched to the matrix vacua and to data from localization \cite{Komatsu:2024bop,Komatsu:2024ydh,Najafian:2026wld}. More general type IIB solutions with the relevant $F(4)$ symmetry have since been constructed \cite{DHoker:2025nid,DHoker:2026eox}. A useful precedent is the holographic treatment of ${\cal N}=1^*$: supergravity RG flows describe the effect of its mass deformation on the bulk geometry \cite{Girardello:1999bd,Pilch:2000fu}, while the Polchinski-Strassler construction identifies polarized brane sources in its vacua \cite{Polchinski:2000uf}. This suggests studying analogous holographic flows driven by the IKKT polarization parameter. Such a study calls for an understanding of how the matrix operators and their dynamics are encoded in bulk fluctuations.

The non-conformal D$p$-brane dualities provide a broader framework for this operator-bulk dictionary \cite{Itzhaki:1998dd,Boonstra:1998mp,Wiseman:2008qa,Kanitscheider:2008kd,Bobev:2025idz}. For the undeformed IKKT model, the tower of BPS multiplets ${\cal B}_n$ of matrix operators is in one-to-one correspondence with the Kaluza-Klein fluctuations of Euclidean type IIB supergravity around the D$(-1)$-instanton background \cite{Morales:2004xc,Ciceri:2025wpb}. The nonlinear dynamics of the lowest Kaluza-Klein multiplet ${\cal B}_2$ are captured by a one-dimensional maximal gauged supergravity \cite{Ciceri:2025maa,Ciceri:2025wpb,Ciceri:2026hxr}, which is embedded as a consistent truncation in the ten-dimensional Euclidean IIB supergravity. 
In this paper, we use this one-dimensional supergravity and its supersymmetric backgrounds to develop a holographic picture of the polarized IKKT model by providing the bulk realization of its lowest BPS multiplet.

In a first step, we analyze the fate of the lowest BPS multiplet ${\cal B}_2$ under the polarization of the IKKT model. Since the deformation modifies the supersymmetry transformations and algebra, maintaining a closed multiplet requires a mixing of the original operators with the Konishi operator, the lowest component of a long supermultiplet of the undeformed model. Furthermore, polarization gives rise to new null relations among the operators which arise as the Ward identities for the broken ${\rm SO}(10)$ symmetry generators. Together with the decoupling of a supersymmetry invariant linear combination of singlets, this results in a known supermultiplet of the superalgebra $F(4)$, carrying 152 bosonic and 144 fermionic states. Once the global symmetries and supersymmetries of the model are taken into account, this finally amounts to 128 bosonic and 128 fermionic degrees of freedom, as expected for a maximally supersymmetric bulk theory.

We then seek to reproduce this structure from the bulk perspective. Since the leading operator for the deformation is part of the ${\cal B}_2$ multiplet, this suggests that the one-dimensional maximal supergravity also encodes the information about the operators of the resulting $F(4)$ multiplet. To make this explicit, we first review the half-BPS solutions of the one-dimensional model that preserve an ${\rm SO}(3)\times{\rm SO}(7)$ bosonic symmetry. We identify the BPS background whose leading term matches the operator defining the IKKT polarization, and present this solution in analytic form. Next, we work out and solve the linearized fluctuation equations around both the undeformed ${\rm SO}(10)$ and polarized ${\rm SO}(3)\times{\rm SO}(7)$ BPS backgrounds. In both cases, we give analytic solutions to the fluctuation equations and identify the independent integration constants. We then determine the action of supersymmetry on these asymptotic data, and find that their multiplet structure precisely reproduces the ${\cal B}_2$ multiplet for the ${\rm SO}(10)$ background, and the previously identified $F(4)$ multiplet for the polarized background. We finally briefly comment on the further steps in the program of holographic renormalization, in particular on the likely need for a Legendre transform in part of the scalar sector.

The rest of this paper is organized as follows. In section \ref{sec:polIKKT}, we review the lowest BPS multiplet ${\cal B}_2$ of the IKKT model and determine how its operators reorganize under polarization into an $F(4)$ multiplet, taking into account the new Ward identities and the decoupled singlet. In section \ref{sec:sugra}, we review the one-dimensional maximal supergravity and identify the ${\rm SO}(3)\times{\rm SO}(7)$-invariant BPS background matching the polarized deformation. Section \ref{sec:asymptotic} derives and solves the linearized bosonic and fermionic fluctuation equations around both the undeformed and polarized backgrounds, identifying their independent asymptotic coefficients. In section \ref{sec:supermultiplets}, we determine the supersymmetry transformations of these coefficients and match the resulting source multiplets to the matrix-model operators. We summarize the results and comment on the remaining steps toward a computation of holographic correlators in section \ref{sec:conclusions}. Two appendices collect conventions and some technical details of the computations.


\section {Polarized IKKT model and BPS multiplets}\label{sec:polIKKT}


In this section, we start by reviewing the structure of the lowest BPS multiplet ${\cal B}_2$ of single trace operators of the undeformed IKKT model. We then review the ${\rm SO}(3)\times{\rm SO}(7)$ invariant polarized IKKT model and study the fate of the ${\cal B}_2$  multiplet under the deformed supersymmetry transformations of the polarized model. Closure of the multiplet requires a particular deformation of the operators. Moreover, in the polarized case, the Ward identity from the broken ${\rm SO}(10)$ generators in the  ${\bf (3,7)}$ induces null relations for 21 of the properly deformed operators from  ${\cal B}_2$. Finally, we identify a supersymmetry invariant singlet operator which after the deformation decouples from the rest. The remaining (deformed) operators of the ${\cal B}_2$ multiplet then form a known irreducible supermultiplet of the simple superalgebra $F(4)$, the symmetry algebra of the polarized IKKT model.

\subsection{IKKT model}
\label{subsec:IKKT}

\subsubsection*{Action and supersymmetry}

Let us briefly review the undeformed IKKT model of \cite{Ishibashi:1996xs}. The Euclidean IKKT action is given by
\begin{equation}
 S_{\text{\tiny{IKKT}}}=\Tr\left[-\frac14[X_a,X_b][X_a,X_b]
 +\frac12\bar\Psi\Gamma^a[X_a,\Psi]\right],
 \label{eq:IKKTaction}
\end{equation}
and supersymmetric under 
\begin{equation}
 \delta_\epsilon X_a=\bar\eps\Gamma_a\Psi,
 \qquad
 \delta_\epsilon \Psi=\frac12\Gamma^{ab}[X_a,X_b]\,\eps\,.
 \label{eq:IKKTsusy}
\end{equation}
Here, $X_a$ and $\Psi^{\alpha}$ are bosonic and fermionic ${\mathfrak su}(N)$-valued matrices, respectively.
Vector indices $a=1, \dots, 10$, are contracted in Euclidean signature, and the $\Gamma_a$ denote the $\mathrm{SO}(10)$ $\Gamma$-matrices. The spinors $\Psi^{\alpha}$, and supersymmetry parameters $\eps^\alpha$, $\alpha=1, \dots, 32$, are complex and satisfy the chirality condition
\bea
\Gamma_{*}\Psi=\Psi\;,\qquad
\Gamma_{*}\,\eps=\eps\;,\qquad
\Gamma_{*}:=-i\,\Gamma_{1}\Gamma_2\ldots\Gamma_{10}
\,.
\label{eq:chirality}
\eea  
We summarize our spinor conventions in Appendix~\ref{app:spinor}. Let us note here, that the charge-conjugated spinors are defined as 
\begin{equation}
    \bar \Psi_\alpha:=\Psi^\beta\mathcal C_{\alpha\beta}
    \,,\label{eq:CCIKKT}
\end{equation}
with the symmetric charge conjugation matrix 
$\mathcal C=-i\Gamma_{10}\Gamma_{*}$, c.f.\ (\ref{eq:CC}). In particular, the 
Hermitian conjugate of $\Psi^\alpha$ does not appear in the (complex) action $S_{\text{\tiny{IKKT}}}$, as is common for spinorial actions in Euclidean signature \cite{Nicolai:1978vc,Krauth:1998xh}. In most formulas, we will suppress the explicit spinor indices.

The equations of motion can be written as
\begin{align}
0 = {\cal E}_a&=[X_b,[X_b,X_a]]- \bar\Psi \Gamma_a \Psi \,,
 \nonumber\\
 0 = {\cal F}&=\Gamma^a[X_a,\Psi]\,.
 \label{eq:IKKTeom}
\end{align}
The supersymmetry algebra closes on-shell, \textit{i.e.}\ upon using the fermionic equations of motion, 
into the bosonic $\mathfrak{su}(N)$ transformations as
\begin{align}
{} [\delta_{\eps_1},\delta_{\eps_2}]\,X_a
 &=\delta_\Lambda X_a\,,
 \nonumber\\
 {} [\delta_{\eps_1},\delta_{\eps_2}]\,\Psi
 &=\delta_\Lambda\Psi-2\,(\bar\eps_2\Gamma^a\eps_1)\Gamma_a\,{\cal F}\,,
 \label{eq:IKKTsusyalgebra}
\end{align}
with $\mathfrak{su}(N)$ parameter $\Lambda=2\,\bar\eps_{2}\Gamma^a\eps_{1}\,X_a$.

The equations of motion and supersymmetry transformations of the undeformed IKKT model are homogeneous under the assignment of weights
\begin{equation}{}
[X^a]=1\,,\quad
[\Psi]=\tfrac32\,,\quad
[\eps]=-\tfrac12
\,,
\label{eq:weight}
\end{equation}
under which the action is of weight $[S_{\text{\tiny{IKKT}}}]=4$\,.
Let us note, that no ${\rm SO}(10)$ rotation can appear in the supersymmetry algebra (\ref{eq:IKKTsusyalgebra}) as a consequence of the scaling symmetry (\ref{eq:weight}) and the spinor chiralities.

\subsubsection*{The lowest BPS multiplet}

Gauge invariant single trace operators\footnote{
In slight abuse of language, but in line with the analogous structures in higher-dimensional Yang-Mills theories, we use the term {\em gauge invariant operators} for $\mathfrak{su}(N)$ invariant operators, despite the fact that all symmetries are global in zero dimensions.
}
 of the model are constructed as traces over products of the fundamental operators $X$ and $\Psi$. The spectrum of these operators and their organisation into short and long supermultiplets has been discussed in some detail in \cite{Morales:2004xc,Ciceri:2025wpb}. Here, we just recall the lowest BPS multiplet ${\cal B}_2$ of such operators, spanned by 
\begin{align}
 {\cal O}_{ab}&={\rm Tr}(X_a X_b)
 -\frac1{10}\,\delta_{ab}{\rm Tr}(X_c X_c)\,,
 \nonumber\\
 {\cal O}_a&={\rm Tr}(X_a\Psi)
 -\frac19\,{\rm Tr}(X^b\Gamma_{ab}\Psi)
 \,,
 \nonumber\\
 {\cal O}_{abc}&={\rm Tr}\bigl(X_a[X_b,X_c]\bigr)
 -\frac18\,{\rm Tr}(\bar\Psi\Gamma_{abc}\Psi)\,.
 \label{eq:OOO}
\end{align}
They transform in the $\mathbf{54}$, $\mathbf{144}$ and $\mathbf{120}$, respectively, of
${\rm SO}(10)$. In particular, ${\cal O}_{ab}$ and ${\cal O}_{abc}$ are symmetric traceless and antisymmetric, respectively, in their indices, while  $\Gamma^a {\cal O}_a=0$. 
With respect to the grading (\ref{eq:weight}), they carry weight $2$, $\frac52$, and $3$, respectively.
The bulk realization of this supermultiplet is the $D=1$ maximal supergravity constructed in \cite{Ciceri:2025maa,Ciceri:2025wpb}.

\begin{figure}[tb]
\begin{center}
\begin{tikzpicture}[x=1.15cm,y=1.15cm]
\node[box,minimum width=34mm] (O54) at (0,1.2)
 {$\mathcal O_{ab}\;\mathbf{54}$};
\node[fbox,minimum width=34mm] (O144) at (0,0)
 {$\mathcal O_a\;\mathbf{144}$};
\node[box,minimum width=34mm] (O120) at (0,-1.2)
 {$\mathcal O_{abc}\;\mathbf{120}$};

\draw[lead] (O54) to (O144);
\draw[lead] (O144) to (O120);
\end{tikzpicture}
\end{center}
\caption{The ${\cal B}_2$ multiplet.}
\label{fig:B2}
\end{figure}
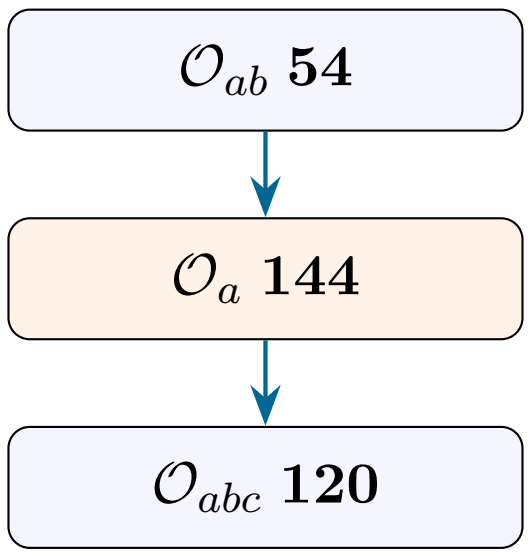

Supersymmetry (\ref{eq:IKKTsusy}) induces the following transformations
\begin{align}
 \delta_\epsilon {\cal O}_{ab}
 &=\frac95\,\bar\eps\Gamma_{(a} {\cal O}_{b)}\,,
 \nonumber\\
 \delta_\epsilon {\cal O}_a
 &=\frac1{18}\left(7\,\Gamma^{bc}\eps\, {\cal O}_{abc}
 -\Gamma_a{}^{bcd}\eps\, {\cal O}_{bcd}\right),
 \nonumber\\
 \delta_\epsilon {\cal O}_{abc}&=0\,,
 \label{eq:OOOsusy}
\end{align}
within the supermultiplet (\ref{eq:OOO}). The last equation requires the fermionic field equations. The nilpotent structure of the multiplet is a consequence of the grading (\ref{eq:weight}) and depicted in Figure~\ref{fig:B2}.
It is straightforward to verify that the commutator of these supersymmetries vanishes 
on the $\mathfrak{su}(N)$ invariant operators.

The representation-theoretic bookkeeping for this multiplet is extrapolated from generic BPS multiplets as
\begin{equation}
 {\cal B}_2=\mathbf{54}_2\ominus\mathbf{144}_{5/2}
 \oplus\mathbf{120}_3\ominus\mathbf{45}_4
 \oplus\mathbf{16}_{9/2}\,.
 \label{eq:B2}
\end{equation}
Subscripts refer to the weight assignment (\ref{eq:weight}).
The first three terms in (\ref{eq:B2}) correspond to the operators (\ref{eq:OOO}) with 174 bosonic and 144 fermionic states. The last two terms account for the global ${\rm SO}(10)$ symmetry and supersymmetry, respectively. In zero dimensions, these symmetries can be used to eliminate scalar and spinor degrees of freedom. Hence the physical count results in $129$ bosonic and $128$ fermionic degrees of freedom. The surprising $+1$ mismatch appears to be a generic feature of the ${\rm SO}(10)$ supermultiplets \cite{Ciceri:2025wpb}.

Let us finally note, that the remaining singlet operator of weight 2
\begin{equation}
{\cal K} = {\rm Tr}(X_a X_a)
\,,
\label{eq:Konishi}
\end{equation}
is not part of the BPS multiplet, but rather the lowest weight state of a long supermultiplet on its own, the analogue of the Konishi multiplet \cite{Bianchi:2001cm} in zero dimensions.

\subsection{Polarized IKKT model}

\subsubsection*{Action and supersymmetry}

The polarized IKKT model is an ${\rm SO}(3)\times {\rm SO}(7)$ invariant deformation of the IKKT matrix model \cite{Bonelli:2002mb}. The bosonic operators are decomposed as
\begin{equation}
\{X^a \} \longrightarrow \{X^{\indI{1}}, X^{\indA{1}}\}\,,\quad \indI{1}=1,2,3\,,\;\;\indA{1}=4, \dots, 10
\,,
\end{equation}
according to the split 
\begin{align}
{\rm SO}(10)\longrightarrow\,&\, {\rm SO}(3)\times {\rm SO}(7)
\,,\nonumber\\
{\bf 10} \longrightarrow\,&\, {\bf (3,1)} \oplus {\bf (1,7)}
\,.
\label{eq:so73}
\end{align}
The spinor remains
irreducible under this breaking
\begin{equation}
{\bf 16} \longrightarrow {\bf (2,8)}
\,.
\label{eq:1682}
\end{equation}
In this notation the polarized IKKT model is given by
\begin{equation}
 S_\Omega=S_{\text{\tiny{IKKT}}}+i\Omega\,
 \Tr\bigg(\frac{1}{6}\epsilon^{\indI{1}\indI{2}\indI{3}}X_{\indI{1}}[X_{\indI{2}},X_{\indI{3}}]
 -\frac{1}{8}\bar\Psi\Gpolar
 \Psi\bigg)
 +\frac{\Omega^2}{64}\,\Tr\big(X_{\indA{1}} X_{\indA{1}}
 +3\,X_{\indI{1}}X_{\indI{1}}
 \big)
 \,,
\label{eq:IKKTpolar}
\end{equation}
with deformation parameter $\Omega$. We have introduced the totally antisymmetric ${\rm SO}(3)$ invariant tensor $\epsilon^{IJK}$, and defined
\begin{equation}
\Gpolar:=\Gamma_{1}\Gamma_2\Gamma_3
=\frac16\,\epsilon^{\indI{1}\indI{2}\indI{3}}\Gamma_{\indI{1}\indI{2}\indI{3}}
\,.\label{eq:Gpolar}
\end{equation}
Comparison with (\ref{eq:OOO}) shows that the leading term of the deformation (\ref{eq:IKKTpolar}) is indeed an operator 
\begin{equation}
{\cal O}_\Omega = \frac16\,\epsilon^{\indI{1}\indI{2}\indI{3}}\,{\cal O}_{\indI{1}\indI{2}\indI{3}}\,,
\label{eq:Odef}
\end{equation}
from the ${\bf 120}$ in the BPS multiplet ${\cal B}_2$\,. 

The deformed equations of motion are
\begin{align}
 0 = {\cal E}_{\indI{1}}&=[X_a,[X_a,X_{\indI{1}}]]- \bar\Psi \Gamma_{\indI{1}} \Psi 
 +\frac{i\Omega}{2}\epsilon_{{\indI{1}}{\indI{2}}{\indI{3}}}\,[X^{\indI{2}},X^{\indI{3}}]
 +\frac{3\Omega^2}{32}X_{\indI{1}}
 \,,
\nonumber\\
0 = {\cal E}_{\indA{1}}&=[X_a,[X_a,X_{\indA{1}}]]- \bar\Psi \Gamma_{\indA{1}} \Psi +\frac{\Omega^2}{32}X_{\indA{1}}\,,
 \nonumber\\
  0 = \cF&=\Gamma^a[X_a,\Psi]-\frac{i\Omega}{4}\Gpolar \Psi\,.
 \label{eq:polarEOM}
\end{align}
The deformation breaks the original ${\rm SO}(10)$ symmetry of the IKKT
model down to ${\rm SO}(3)\times{\rm SO}(7)$.  Accordingly, variation of
the action \eqref{eq:IKKTpolar} under the broken symmetry generators in the
${\bf(3,7)}$ gives rise to the Ward identity
\begin{equation}
 \Tr\!\left(\!X_{\indA{1}}{\cal E}_{\indI{1}}-X_{\indI{1}}{\cal E}_{\indA{1}}
 +\frac12\,\bar{\cF}\,\Gamma_{{\indI{1}}{\indA{1}}}\Psi\right)
 =
 i\Omega\left(
 \frac{ 1}2\,\epsilon_{\indI{1}}{}^{{\indI{2}}{\indI{3}}}{\cal O}_{{\indI{2}}{\indI{3}}{\indA{1}}}
 -\frac{i\Omega}{16}\,{\cal O}_{{\indI{1}}{\indA{1}}}\right)
 \,=:\, 
 i\Omega\,{\cal N}_{{\indI{1}}{\indA{1}}}\,,
 \label{eq:Ward}
\end{equation}
at order $\Omega$.
Thus, for $\Omega\neq0$, ${\cal N}_{{\indI{1}}{\indA{1}}}$ defines a non-trivial
on-shell null relation among the operators \eqref{eq:OOO}.  This relation
disappears in the limit $\Omega\to0$, in which the full ${\rm SO}(10)$
symmetry is restored and \eqref{eq:Ward} reduces to the corresponding
Noether identity.

The polarized IKKT model (\ref{eq:IKKTpolar}) is invariant under the deformed supersymmetry transformations
\begin{equation}
 \delta_\epsilon X^a=\bar\eps \Gamma^a \Psi\,,
 \qquad
 \delta_\epsilon\Psi=\frac12\Gamma^{ab}[X_a,X_b]\,\eps
 -\frac{i\Omega}{8}\,\big(\Gamma^a\Gpolar+2\Gpolar\Gamma^a\big) X_a\eps\,,
 \label{eq:polarSUSY}
\end{equation}
thus preserving the full number of supersymmetries of the undeformed IKKT model.
The equations of motion (\ref{eq:polarEOM})
are mapped into each other under the supersymmetry (\ref{eq:polarSUSY}).

\subsubsection*{Supersymmetry algebra}

The algebra of deformed supersymmetry transformations (\ref{eq:polarSUSY}) closes on-shell according to
\begin{align}
 {} [\delta_{\eps_1},\delta_{\eps_2}]\,X_a
 &=\delta_\Lambda X_a
 +\Theta_{a}{}^bX_b 
 \,,
 \nonumber\\
{} [\delta_{\eps_1},\delta_{\eps_2}]\,\Psi
 &=\delta_\Lambda\Psi
 +\frac14\Theta^{ab}\Gamma_{ab}\,\Psi
 -2\,(\bar\eps_2\Gamma^a\eps_1)\Gamma_a\,{\cal F}
 \,,
 \label{eq:IKKTpolarsusyalgebra}
\end{align}
with $\mathfrak{su}(N)$ parameter $\Lambda=2\,\bar\eps_{2}\Gamma^a\eps_{1}\,X_a$,
the fermionic field equations ${\cF}$, and 
an ${\rm SO}(3)\times{\rm SO}(7)$ rotation with parameter $\Theta^{ab}$ given by
\begin{equation}
 \Theta^{\indI{1}\indI{2}}=\frac{3i\Omega}{4}\bar\eps_{[2}\Gpolar\Gamma^{\indI{1}\indI{2}}\eps_{1]}\,,\qquad
 \Theta^{\indA{1}\indA{2}}=-\frac{i\Omega}{4}\bar\eps_{[2}\Gpolar\Gamma^{\indA{1}\indA{2}}\eps_{1]}\,,\qquad
 \Theta^{\indI{1}\indA{1}}=0\,.
 \label{eq:Thetaexplicit}
\end{equation}
Equations \eqref{eq:IKKTpolarsusyalgebra} are the
polarized deformation of \eqref{eq:IKKTsusyalgebra}.  In contrast to the unpolarized model, the
on-shell supersymmetry algebra does not become abelian when restricted to $\mathfrak{su}(N)$ singlet operators,
but rather yields the exceptional simple Lie superalgebra $F(4)$\footnote{This is the complex form of the Lie algebra $F(4,\mathbb C)$, in accordance with the fact that in Euclidean signature the operators $X^a$ and $\Psi$ are complex.} \cite{Komatsu:2024ydh,Martina:2025kwc,DHoker:2025nid}
\begin{equation}
 \{Q_\alpha,Q_\beta\}=\frac{3i\Omega}{8}
 \left[
 ({\cal C}\Gpolar\Gamma^{\indI{1}\indI{2}})_{\alpha\beta}M_{\indI{1}\indI{2}}
 -\frac13({\cal C}\Gpolar\Gamma^{\indA{1}\indA{2}})_{\alpha\beta}M_{\indA{1}\indA{2}}\right]
 .
 \label{eq:F4algebra}
\end{equation}
Its bosonic subalgebra is
$\mathfrak{so}(3)\oplus\mathfrak{so}(7)$ with generators $M_{\indI{1}\indI{2}}$ and $M_{\indA{1}\indA{2}}$ normalized as
\begin{equation}
 [M_{ab},M_{cd}]=\delta_{bc}M_{ad}-\delta_{ac}M_{bd}
 -\delta_{bd}M_{ac}+\delta_{ad}M_{bc}\,,
 \label{eq:uniformso}
\end{equation}
while the odd generators transform as ${\bf(2,8)}$.

\subsection{$F(4)$ supersymmetry and BPS multiplets}
\label{subsec:F4}

With (\ref{eq:polarSUSY}), it is straightforward to compute the supersymmetry transformation of single-trace operators over products of fundamental operators. However, the deformed supersymmetry transformations necessitate a new analysis of the resulting multiplet structures.
In particular, as the deformation is triggered by a dimensionful parameter 
\begin{equation}
[\Omega]=1\,,
\label{eq:omega}
\end{equation}
the resulting new multiplets will no longer retain the nilpotent structure of the IKKT multiplets (c.f.\ Figure \ref{fig:B2}).\footnote{Note that the $\Omega$ deformation breaks the scaling symmetry of the equations of motion \eqref{eq:weight}.} We postpone a systematic treatment to future work; here, we focus on  
the fate of the lowest BPS supermultiplet ${\cal B}_2$ (\ref{eq:B2}) of the IKKT model under the deformation. In principle, the deformed supersymmetry transformations (\ref{eq:polarSUSY}) of the fundamental operators may --- and indeed do --- induce a leakage of the former BPS supermultiplet into the long Konishi multiplet. However, we will see that suitable redefinition of the operators, as well as the new null relation (\ref{eq:Ward}), allow to maintain the structure of a BPS multiplet under the new supersymmetry algebra $F(4)$.

As a first step, we decompose the field content of the ${\cal B}_2$ supermultiplet under ${\rm SO}(3)\times {\rm SO}(7)$
\begin{align}
 \mathbf{54}&\longrightarrow\rep{5}{1}\oplus\rep{1}{27}
 \oplus\rep{3}{7}\oplus\rep{1}{1}\,,
 \nonumber\\
 \mathbf{144}&\longrightarrow\rep{4}{8}\oplus\rep{2}{48}
 \oplus\rep{2}{8}\,,
 \nonumber\\
 \mathbf{120}&\longrightarrow\rep{1}{35}
 \oplus\rep{3}{21} \oplus\rep{3}{7} \oplus \rep{1}{1}
 \,.
 \label{eq:branchB2}
\end{align}
We then study the action of supersymmetry successively. Let us start from the operators ${\cal O}_{ab}$ (\ref{eq:OOO})
from the ${\bf 54}$, which we denote as
\begin{equation}
{\bf 54}\,: \quad
\left\{
\begin{array}{l}
{\cal B}_{\indI{1}\indI{2}} = {\cal O}_{\indI{1}\indI{2}}-\frac13\,\delta_{\indI{1}\indI{2}}\,{\cal O}_{\indI{3}\indI{3}}\,,\quad
{\cal B}_{\indA{1}\indA{2}} = {\cal O}_{\indA{1}\indA{2}}-\frac17\,\delta_{\indA{1}\indA{2}}\,{\cal O}_{\indA{3}\indA{3}}\,,\\
{\cal B}_{\indI{1}\indA{1}} = {\cal O}_{\indI{1}\indA{1}}\,,\qquad
{\cal B}={\cal O}_{\indA{1}\indA{1}}
\,,
\end{array}
\right.
\label{eq:break54}
\end{equation}
in accordance with the first line of (\ref{eq:branchB2}). Since these operators are exclusively build from the bosonic fundamental operators $X^a$, their supersymmetry transformation is unchanged and still given by the first line of (\ref{eq:OOOsusy}). To spell this out for the different components of (\ref{eq:break54}), we decompose the operators ${\cal O}_a$ (\ref{eq:OOO})
from the ${\bf 144}$ according to the second line of (\ref{eq:branchB2}) into
\begin{equation}
{\bf 144}\,:\quad
\Big\{\;\;
\Upsilon_{\indI{1}} = \mathbb{P}_{\indI{1}}{}^{\indI{2}}{\cal O}_{\indI{2}}
\,,\qquad
\Upsilon_{\indA{1}}= \mathbb{P}_{\indA{1}}{}^{\indA{2}}{\cal O}_{\indA{2}} 
\,,\qquad
\Upsilon = \Gamma^{\indA{1}} {\cal O}_{\indA{1}}
\,,
\label{eq:break144}
\end{equation}
where we have introduced the spinor projectors
\begin{equation}
\mathbb{P}_{\indI{1}}{}^{\indI{2}}
:= \delta_{\indI{1}}{}^{\indI{2}} - \frac13\,\Gamma_{\indI{1}}{}\Gamma^{\indI{2}}
\,,\qquad
\mathbb{P}_{\indA{1}}{}^{\indA{2}}
:= \delta_{\indA{1}}{}^{\indA{2}} - \frac17\,\Gamma_{\indA{1}}{}\Gamma^{\indA{2}}
\,,
\label{eq:proj37}
\end{equation}
to implement the projection onto the ${\rm SO}(3)\times{\rm SO}(7)$ irreducible spinor representations. In particular, we have $\Gamma_{\indI{1}} \Upsilon^{\indI{1}} =0=\Gamma_{\indA{1}} \Upsilon^{\indA{1}}$.
Supersymmetry then transforms the operators (\ref{eq:break54}) into (\ref{eq:break144}) as
\begin{align}
 \delta_\eps {\cal B}_{\indI{1}\indI{2}}&=\frac95\,\bar\eps\,
 \Gamma_{(\indI{1}} \Upsilon_{\indI{2})}\,,
\qquad
 \delta_\eps {\cal B}_{\indA{1}\indA{2}}=\frac95\,\bar\eps\,
 \Gamma_{(\indA{1}}  \Upsilon_{\indA{2})}\,,
  \nonumber\\
\delta_\eps {\cal B}_{\indI{1}\indA{1}}&=\frac9{10}\,\bar\eps\,
 (\Gamma_{\indI{1}}\Upsilon_{\indA{1}}
 +\Gamma_{\indA{1}}\Upsilon_{\indI{1}})
 +\frac37\,\bar\eps\,\Gamma_{{\indI{1}}{\indA{1}}}\Upsilon\,,
 \qquad
 \delta_\eps {\cal B} =\frac95\,\bar\eps\Upsilon\,,
 \label{eq:delta54}
\end{align}
independently of the deformation parameter $\Omega$.
Next, applying supersymmetry to the fermionic operators (\ref{eq:break144}) maps these into representations from the ${\bf 120}$, c.f.\ (\ref{eq:OOOsusy}), and due to the presence of the weighted deformation parameter $\Omega$ (\ref{eq:omega}) also back into representations from the ${\bf 54}$. To make this explicit, we denote the operators in the ${\bf 120}$ as 
\begin{equation}
{\bf 120}\,: \quad
\left\{
\begin{array}{l}
{\cal A}_{\indA{1}\indA{2}\indA{3}} = {\cal O}_{\indA{1}\indA{2}\indA{3}}\,,\qquad
{\cal A}_{\indI{1}\indA{1}\indA{2}} = {\cal O}_{\indI{1}\indA{1}\indA{2}}\,,
\\
{\cal N}_{\indI{1}\indA{1}} = 
\frac12\,\epsilon_{\indI{1}}{}^{\indI{2}\indI{3}}\,{\cal O}_{\indI{2}\indI{3}\indA{1}}
-\frac{i\Omega}{16}\,{\cal B}_{IA}
\,,
\qquad
{\cal A} = 
\frac16\,\epsilon^{\indI{1}\indI{2}\indI{3}}\,{\cal O}_{\indI{1}\indI{2}\indI{3}}
+\frac{3i\Omega}{28}\,{\cal B}-\frac{i\Omega}{20}\,{\cal K}\,,
\end{array}
\right.
\label{eq:break120}
\end{equation}
in accordance with the third line of (\ref{eq:branchB2}). Here, we have included an $\Omega$-dependent deformation of the last two representations in the ${\bf (3,7)}$ and the ${\bf (1,1)}$, respectively, in order to simplify the resulting transformations. Note, that the corrected ${\bf (3,7)}$ reproduces the null relation (\ref{eq:Ward}) and thus vanishes on-shell if $\Omega\not=0$. The corrected singlet ${\bf (1,1)}$ includes a contribution from the Konishi singlet (\ref{eq:Konishi}) which was part of a different multiplet in the undeformed IKKT model.
The supersymmetry transformation of the fermionic operators (\ref{eq:break144}) then reads
\begin{align}
 \delta_\eps \Upsilon_{\indI{1}}={}&\mathbb{P}_{\indI{1}}{}^{\indI{2}}\,
 \Bigl(
  \frac59\,{\cal A}_{{\indI{2}}{\indA{1}\indA{2}}}\,\Gamma^{\indA{1}\indA{2}}
  -\frac{5i\Om}{24}\left(
   2{\cal B}_{\indI{2}\indI{3}}\,\Gamma^{\indI{3}}
   -{\cal B}_{\indI{2}\indA{1}}\,\Gamma^{\indA{1}}
  \right)\Gpolar
  +\frac{10}{9}\,{\cal N}_{\indI{2}\indA{1}}\,\Gamma^{\indA{1}}\Gpolar
 \Bigr)\,\eps\,,
\nonumber\\[1ex]
 \delta_\eps \Upsilon_{\indA{1}}={}&\mathbb{P}_{\indA{1}}{}^{\indA{2}}\,
 \Bigl(
  -\frac{10}{9}\,{\cal A}_{\indI{1}\indA{2}\indA{3}}\Gamma^{\indI{1}\indA{3}}
  +\frac59\,{\cal A}_{\indA{2}\indA{3}\indA{4}}\Gamma^{\indA{3}\indA{4}}
  -\frac{5i\Om}{72}\left(
  5{\cal B}_{\indI{1}\indA{2}}\Gamma^{\indI{1}}
   -2{\cal B}_{\indA{2}\indA{3}}\Gamma^{\indA{3}}
  \right)\Gpolar
 +\frac{10}{9}\,{\cal N}_{\indI{1}\indA{2}}\,\Gpolar\Gamma^{\indI{1}}
 \Bigr])\,\eps\,,
 \nonumber\\[1ex]
\delta_\eps \Upsilon={}&\Bigl(
  -\frac1{18}\,{\cal A}_{\indI{1}\indA{1}\indA{2}}\Gamma^{\indI{1}\indA{1}\indA{2}}
  +\frac16\,{\cal A}_{\indA{1}\indA{2}\indA{3}}\Gamma^{\indA{1}\indA{2}\indA{3}}
  -\frac73{\cal A}\,\Gpolar
  - \frac{5i\Om}{48}\,{\cal B}_{\indI{1}\indA{1}} \,\Gpolar\Gamma^{\indI{1}\indA{1}}
  -\frac{11}{9}\,{\cal N}_{\indI{1}\indA{1}}\,\Gpolar\Gamma^{\indI{1}\indA{1}}
 \Bigr)\,\eps\,.
 \label{eq:delta144}
\end{align}
 They encode the $\Omega$ deformation through both explicit $\Omega$-dependent terms and the operator redefinitions (\ref{eq:break120}).

\begin{figure}[tb]
\begin{center}
\begin{tikzpicture}[x=1.15cm,y=1.15cm]
\node[box] (B) at (-6,3) {${\cal B}_{\indI{1}\indI{2}}\;\rep{5}{1}$};
\node[box] (C) at (-2,3) {${\cal B}_{\indA{1}\indA{2}}\;\rep{1}{27}$};
\node[box] (R) at (2,3) {${\cal B}_{\indI{1}\indA{1}}\;\rep{3}{7}$};
\node[box] (H) at (6,3) {${\cal B}\;\rep{1}{1}$};
\node[fbox] (S) at (-4,0) {$\Upsilon^{\indI{1}}\;\rep{4}{8}$};
\node[fbox] (r) at (0,0) {$\Upsilon^{\indA{1}}\;\rep{2}{48}$};
\node[fbox] (T) at (4,0) {$\Upsilon\;\rep{2}{8}$};
\node[nullbox] (Y) at (-6,-3) {${\cal N}_{\indI{1}\indA{1}}\;\rep{3}{7}$ {\scriptsize(EOM-null)}};
\node[box] (D) at (-2,-3) {${\cal A}_{\indA{1}\indA{2}\indA{3}}\;\rep{1}{35}$};
\node[box] (A) at (2,-3) {${\cal A}_{\indI{1}\indA{1}\indA{2}}\;\rep{3}{21}$};
\node[box] (Z) at (6,-3) {${\cal A}\;\rep{1}{1}$};

\foreach \u/\v in {B/S,C/r,R/S,R/r,R/T}{\pairarrows{\u}{\v}{6}}
\draw[lead] (H) to (T);
\foreach \u/\v in {S/A,r/A,T/A,T/Z}{\pairarrows{\u}{\v}{6}}
\pairarrows{r}{D}{6}
\pairarrows{T}{D}{6}
\foreach \u/\v in {S/Y,r/Y,T/Y}{\draw[null,bend right=5] (\u) to (\v);}
\end{tikzpicture}
\end{center}
\caption{The lowest BPS multiplet for the polarized IKKT model. The arrows denote the transformation under supersymmetry. The blue arrows are already present in the undeformed ${\rm SO}(10)$ invariant model, the dashed arrows come with the deformation in order $\Omega$. The lower left block corresponds to the null relations (\ref{eq:Ward}) present in the polarized model.}
\label{fig:Bpolar}
\end{figure}
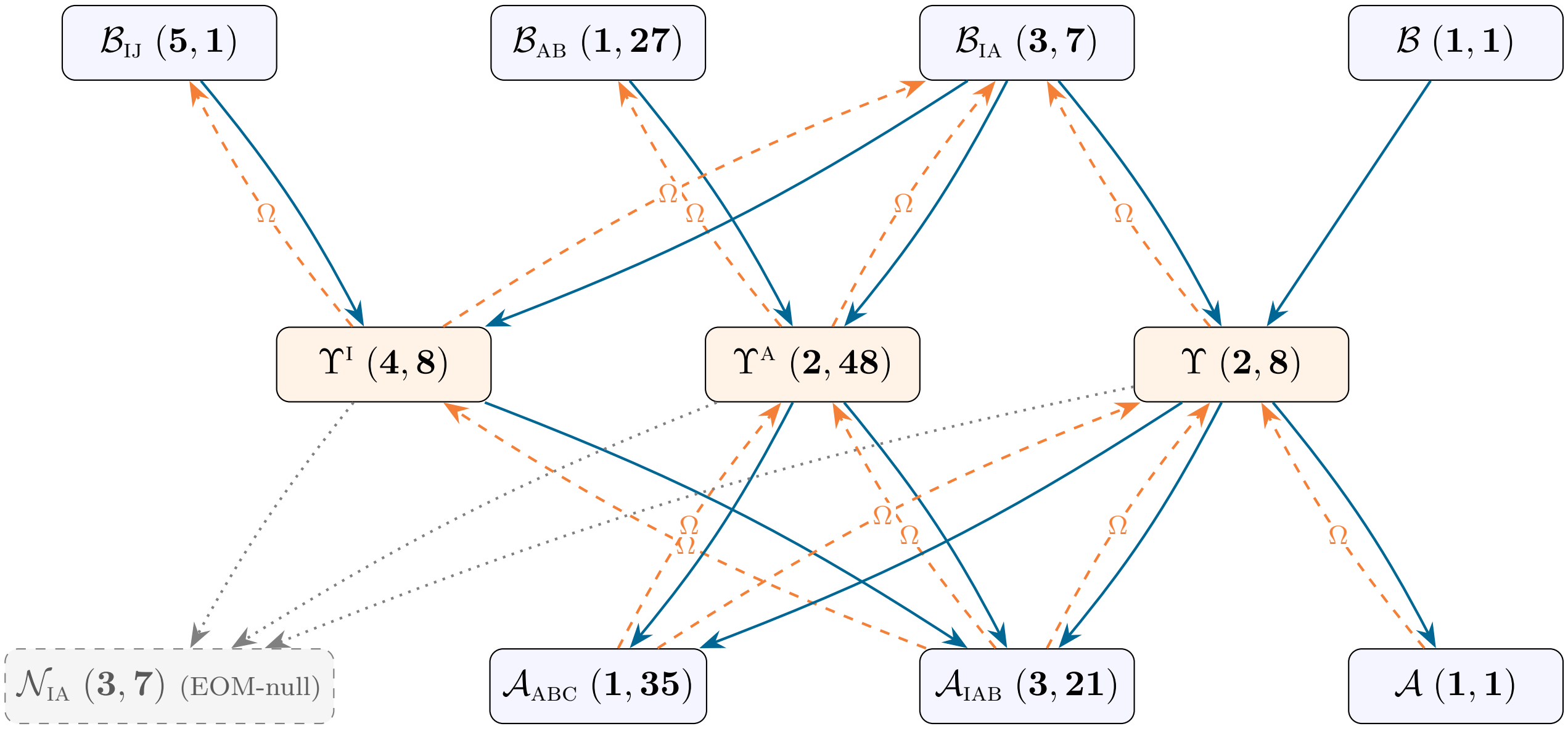

As a final step, it remains to compute the supersymmetry variation of the operators (\ref{eq:break120}) in order to verify that the set of operators (\ref{eq:break54}), (\ref{eq:break144}), (\ref{eq:break120}), form a closed supermultiplet. For the ${\cal N}_{\indI{1}\indA{1}}$, we recall that they can be expressed as linear combinations of the equations of motion (\ref{eq:Ward}). Supersymmetry covariance of the equations of motion therefore ensures that their variations vanish on-shell. For the variation of the remaining operators, we find
\begin{align}
 \delta_\eps {\cal A}_{\indA{1}\indA{2}\indA{3}}={}&
 -\frac{27i\Omega}{160}\,\bar\eps \Gpolar\Gamma_{[\indA{1}\indA{2}}\,\Upsilon_{\indA{3}]}
 -\frac{9i\Omega}{112}\,\bar\eps \Gpolar\Gamma_{\indA{1}\indA{2}\indA{3}}\,\Upsilon
 \,,
 \nonumber\\
\delta_\eps {\cal A}_{\indI{1}\indA{1}\indA{2}}={}&i\Om\left(
 \frac{27}{80}\,\bar\eps \Gpolar \Gamma_{\indI{1}[\indA{1}}\,\Upsilon_{\indA{2}]}
 -\frac{9}{32}\,\bar\eps\, \Gpolar\Gamma_{\indA{1}\indA{2}}\,\Upsilon_{\indI{1}}
 -\frac3{112}\,\bar\eps \Gpolar\Gamma_{\indI{1}\indA{1}\indA{2}}\,\Upsilon\right),
 \nonumber\\
  \delta_\eps\,{\cal A}
 ={}&\frac{9i\Omega}{112}\,\bar\eps \Upsilon
 \,,
 \label{eq:delta120}
\end{align}
up to field equations. For this computation, one has to take into account the deformed supersymmetry transformation rules (\ref{eq:polarSUSY}), the deformed operators (\ref{eq:break120}), as well as the deformed field equations (\ref{eq:polarEOM}). The operators (\ref{eq:break120}) thus close on-shell into the previous operators of the multiplet. The resulting multiplet structure is depicted in Figure~\ref{fig:Bpolar}. 
Finally, we note that we can define a linear combination of the two ${\bf (1,1)}$ singlet operators ${\cal A}$ and ${\cal B}$ whose supersymmetry variation vanishes on-shell 
\begin{equation}
{\cal S} = {\cal A} -\frac{5i\Omega}{112}\, {\cal B}
\,,\qquad
\delta_\epsilon{\cal S} = 0
\,.
\label{eq:singlet}
\end{equation} 
This combination of operators therefore decouples from the supermultiplet. We are then left with an on-shell supermultiplet spanned by the operators
\begin{equation}
{\cal M}_{\rm polar}\;=\;
\Big\{
{\cal B}_{\indI{1}\indI{2}} ,
{\cal B}_{\indA{1}\indA{2}},
{\cal B}_{\indI{1}\indA{1}},
\Upsilon^{\indI{1}},
\Upsilon^{\indA{1}},
\Upsilon,
{\cal A}_{\indA{1}\indA{2}\indA{3}} ,
{\cal A}_{\indI{1}\indA{1}\indA{2}} ,
{\cal A}
\Big\}
\,,
\label{eq:F42d}
\end{equation}
combining 152 bosonic and 144 fermionic states.\footnote{Note that the limit $\Omega\to0$ of this multiplet structure is singular. The null relations \eqref{eq:Ward} disappear, and the unpolarized ${\cal B}_2$ is only recovered by adding back the 21 operators ${\cal N}_{\indI{1}\indA{1}}$ and the singlet ${\cal B}$. This is reminiscent of the localization computation of \cite{Komatsu:2024ydh}, in which the partition function of the polarized IKKT model does not trivially reduce to that of the undeformed model as $\Omega\to0$.}
 This is the atypical supermultiplet $L(2\delta)$ of the superalgebra $F(4)$ \cite{Martirosyan:2014}.

In analogy to the unpolarized multiplet (\ref{eq:B2}), we can denote the full structure obtained from decomposing ${\cal B}_2$ as
\begin{align}
 &
  \rep{5}{1}\oplus\rep{1}{27}
 \oplus\rep{3}{7}
\ominus\rep{4}{8}\ominus\rep{2}{48}
\ominus\rep{2}{8}
  \oplus\rep{1}{35}
 \oplus\rep{3}{21} \oplus \rep{1}{1}
\nonumber\\
& \ominus\rep{3}{1} \ominus\rep{1}{21}
 \oplus\rep{2}{8}\,.
 \label{eq:Bpolar}
\end{align}
The first line carries the 152 bosonic and 144 fermionic operators of (\ref{eq:F42d}). The second line captures the global symmetries which in zero dimensions can be used to eliminate degrees of freedom. These are the 24 bosonic generators of $\mathfrak{so}(3)\oplus\mathfrak{so}(7)$ and the 16 fermionic generators of global supersymmetry. The total count thus amounts to 128 bosonic and 128 fermionic degrees of freedom, the extra bosonic singlet we had encountered in (\ref{eq:B2}) has consistently decoupled in the form of (\ref{eq:singlet}).

In the following, we will study the bulk realization of this multiplet within the one-dimensional maximal gauged supergravity. In particular, we will recover this multiplet in the asymptotic expansion of the supergravity fields around the ${\rm SO}(3)\times{\rm SO}(7)$ BPS domain wall solution.


\section {Supergravity dual}\label{sec:sugra}

In this section, we review the $D=1$ maximal gauged supergravity of \cite{Ciceri:2025maa,Ciceri:2025wpb,Ciceri:2026hxr} which yields the bulk dual realizing the lowest BPS multiplet of the undeformed and the polarized IKKT matrix model, respectively, upon expansion around different BPS backgrounds. We discuss its general BPS solutions preserving an ${\rm SO}(3)\times{\rm SO}(7)$ bosonic symmetry, and their uplift to Euclidean IIB supergravity. We refer to the original papers for more details.


\subsection{D=1 supergravity}

The maximally supersymmetric gauged model in one dimension furnishes a bulk description of the lowest half-BPS multiplet of gauge-invariant single-trace operators in the IKKT model.  Its degrees of freedom are
\begin{equation}
 {\rm bosonic:}\; \{ {\rm e},\phi,A_{[ij]},T_{(ij)},a_{[ijk]}\},
 \qquad
 {\rm fermionic:}\; \{\psi^\alpha,\lambda^\alpha,\chi_a^\alpha\},
\end{equation}
where $i,j=1,\ldots,10$, $a=1,\ldots,10$, and spinor indices $\alpha=1, \dots 32$.  Here ${\rm e}$ is the einbein, $\phi$ is the dilaton, $A_{ij}=-A_{ji}$ comprises 45 (one-dimensional) gauge fields, $T_{ij}=V_i{}^aV_j{}_a$ is a symmetric ${\rm SL}(10)$ representative with 54 independent scalar components, and $a_{ijk}=a_{[ijk]}$ describes 120 axion fields.  The fermionic fields are the gravitino $\psi$, dilatino $\lambda$, and matter fields $\chi_a$ (with $\Gamma^a\chi_a=0$). Charge-conjugated spinors are defined as in \eqref{eq:CCIKKT}. In contrast to the IKKT fermions \eqref{eq:chirality}, the supergravity fermions are not chiral, but carry all 32 complex components. Two useful ${\rm SO}(10)$-covariant scalar combinations are
\begin{equation}
 \mathcal T^{ab}=\delta^{ij}V_i{}^aV_j{}^b,
 \qquad
 a_{abc}=(V^{-1})_a{}^i(V^{-1})_b{}^j(V^{-1})_c{}^k a_{ijk}.
 \label{eq:axionSO-new}
\end{equation}

Keeping terms through quadratic order in fermions, the supergravity Lagrangian can be organized according to powers of the gauge coupling $\mathrm g$
\begin{equation}
 \mathcal L_{\rm SUGRA}=\mathcal L_0+\mathrm g\,\mathcal L_{\rm top}
 +\mathrm g\,\mathcal L_{\rm Yukawa}+\mathrm g^2\mathcal L_{\rm pot}.
 \label{eq:LSUGRA-new}
\end{equation}
Gauge fields and the coupling constant $\mathrm g$ enter the first term $\mathcal L_0$  only through covariant quantities.  Explicitly, this term combines all kinetic terms with the couplings of scalar currents to fermion currents
\begin{align}
\mathcal L_0={}&10\,{\rm e}^{-1}\dot\phi^2-{\rm e}^{-1}P^{ab}P_{ab}
-\frac1{12}{\rm e}^{-1}e^{-2\phi}p_{abc}p^{abc}
+20\bar\lambda\mathcal D_t\lambda+2\bar\chi^a\mathcal D_t\chi_a \nonumber\\
&-20{\rm e}^{-1}\bar\psi\Gamma_*\lambda\dot\phi
+2{\rm e}^{-1}\bar\psi\Gamma_b\chi_a p^{ab}
-\frac12{\rm e}^{-1}e^{-\phi}\bar\chi_a\Gamma_{bc}\psi \,p^{abc}\nonumber\\
&-\frac1{12}e^{-\phi}\bar\chi^a\Gamma_{bcd}\chi_a \,p^{bcd}
-e^{-\phi}\bar\chi_a\Gamma_b\chi_c \,p^{abc}
-\frac16{\rm e}^{-1}e^{-\phi}\bar\lambda\Gamma_{abc}\Gamma_*\psi \,p^{abc}\nonumber\\
&-e^{-\phi}\bar\lambda\Gamma_{abc}\lambda \,p^{abc}
-e^{-\phi}\bar\chi_c\Gamma_{ab}\Gamma_*\lambda \,p^{abc}\,.
\label{eq:L0-new}
\end{align}
The covariant scalar currents appearing here are defined as
\begin{align}
 (V^{-1})^a{}^iD_tV_i{}^b
 &=(V^{-1})^a{}^i\bigl(\dot V_i{}^b-\mathrm g A_{ji}V_j{}^b\bigr)
 =J^{ab}=Q^{[ab]}+P^{(ab)}\,,\\
 (V^{-1})_a{}^i(V^{-1})_b{}^j(V^{-1})_c{}^kD_ta_{ijk}
 &=(V^{-1})_a{}^i(V^{-1})_b{}^j(V^{-1})_c{}^k
 \bigl(\dot a_{ijk}-3\mathrm g A_{l[i}a_{jk]l}\bigr)=p_{abc}\,,
 \nonumber
\end{align}
with ${\rm SO}(10)$ covariant derivatives $D_t$ carrying the gauge fields $A_{ij}$.
A dot denotes differentiation with respect to $t$.  The covariant spinor derivatives include the composite ${\rm SO}(10)$ connection $Q^{ab}$ as
\begin{align}
 \mathcal D_t\psi&=\dot\psi+\frac14\,Q^{ab}\Gamma_{ab}\psi\,,
 \nonumber\\
 \mathcal D_t\chi_a&=\dot\chi_a+\frac14\,Q^{bc}\Gamma_{bc}\chi_a+Q^{ab}\chi_b\,,
\end{align}
with the same connection prescription for $\lambda$.  Since the theory is one-dimensional, neither an Einstein--Hilbert term, nor a Yang-Mills term, nor a Rarita--Schwinger kinetic term exists.  Accordingly, ${\rm e}$, $\psi$, and $A_{ij}$ couple algebraically, inducing constraints rather than equations of motion.

The one-derivative axionic contribution $\mathcal L_{\rm top}$ is the analogue of a topological term in one dimension, 
\begin{align}
 \mathcal L_{\rm top}={}&\frac{\mathrm i}{1152}
 \varepsilon_{a_1a_2a_3a_4a_5a_6a_7a_8a_9a_{10}}
 \mathcal T_{bc}a^{ba_1a_2}a^{ca_3a_4}a^{a_5a_6a_7}p^{a_8a_9a_{10}}\,,
 \label{eq:Ltop-new}
\end{align}
and in particular independent of the einbein ${\rm e}$.
The fermion mass and Yukawa structures may be collected as
\begin{equation}
 \mathcal L_{\rm Yukawa}=-40\bar\lambda B\psi-4\bar\chi_aC^a\psi
 +{\rm e}\bar\lambda E\lambda+{\rm e}\bar\lambda E^a\chi_a
 +{\rm e}\chi_aE^{ab}\chi_b\,.
 \label{eq:YukawaAnsatz-new}
\end{equation}
Their scalar-dependent coefficients are
\begin{align}
B={}&\frac b{10}+\frac3{80}b^{abc}\Gamma_{abc}+\frac1{160}b^{abcd}\Gamma_{abcd}\,,\nonumber\\
C^a={}&-\frac12b^{ab}\Gamma_b\Gamma_*
-\left(\frac7{80}b_{bcd}-\frac16b_{b,cd}\right)\widetilde\Gamma^{ab,cd}\Gamma_*
+\frac3{40}b_{bcde}\widetilde\Gamma^{ab,cde}\Gamma_*\,,\nonumber\\
E={}&-21b\Gamma_*-7b^{abc}\Gamma_{abc}\Gamma_*
-\frac98b^{abcd}\Gamma_{abcd}\Gamma_*\,,\nonumber\\
E^a={}&-16b^{ab}\Gamma_b+(3b^{abc}-4b^{a,bc})\Gamma_{bc}
-2b^{abcd}\Gamma_{bcd}\,,\nonumber\\
E^{ab}={}&-\left(4b^{ab}-\frac1{10}\delta^{ab}b\right)\Gamma_*
-\frac14\delta^{ab}b^{cde}\Gamma_{cde}\Gamma_*
+\frac13(3b^{acb}-4b^{c,ab})\Gamma_c\Gamma_*\nonumber\\
&+\frac1{16}\delta^{ab}b^{cdef}\Gamma_{cdef}\Gamma_*
-\frac32b^{abcd}\Gamma_{cd}\Gamma_*\,,
\label{eq:YukawaAll-new}
\end{align}
given in terms of the irreducible ${\rm SO}(10)$ objects 
\begin{align}
b&=e^{4\phi}\mathcal T^a{}_a\,,\nonumber\\
b^{ab}&=e^{4\phi}\big(\mathcal T^{ab}-\frac1{10}\delta^{ab}\mathcal T^c{}_c\big)\,,\nonumber\\
b^{abc}&=e^{3\phi}\mathcal T^{d[a}a^{bc]}{}_d\,,\nonumber\\
b^{a,bc}&=e^{3\phi}\left(\mathcal T^{da}a^{bc}{}_d-\mathcal T^{d[b}a^{c]a}{}_d\right),\nonumber\\
b^{abcd}&=e^{2\phi}\mathcal T^{ef}a_e{}^{[ab}a^{cd]}{}_f\,.
\label{eq:b-objs}
\end{align}
and the gamma-matrix combinations
\begin{equation}
 \widetilde\Gamma^{a_1a_2,\,a_3\ldots a_n}
 :=\delta^{a_1a_2}\Gamma^{a_3\ldots a_n}
 -\frac1{11-n}\Gamma^{a_1\ldots a_n}\,,
 \label{eq:Gammacombi-new}
\end{equation}
satisfying $\Gamma_{a_1}\widetilde\Gamma^{a_1[a_2,\,a_3\ldots a_n]}=0$.

The remaining bosonic interactions are encoded in the scalar potential
\begin{align}
 \mathcal L_{\rm pot}=-{\rm e}\,V_{\rm pot}=-{\rm e}\,\biggl(
 b^{ab}b_{ab}-\frac25b^2-\frac14b^{abc}b_{abc}
 +\frac29b^{a,bc}b_{a,bc}+\frac3{16}b^{abcd}b_{abcd}\biggr).
 \label{eq:Lpot-new}
\end{align}
In terms of the dressed scalars (\ref{eq:axionSO-new}), the potential becomes 
\begin{align}
V_{\rm pot}={}&\frac12e^{8\phi}\left(2\mathcal T^{ab}\mathcal T_{ab}-(\mathcal T^a{}_a)^2\right)
-\frac14e^{6\phi}\left(2\mathcal T^{ac}\mathcal T^{bd}a_{abe}a_{cd}{}^e
-\mathcal T^{ac}\mathcal T^b{}_c a_{ade}a_b{}^{de}\right)\nonumber\\
&+\frac3{16}e^{4\phi}\mathcal T^{ef}\mathcal T^{gh}
a_e{}^{[ab}a^{cd]}{}_f a_{gab}a_{cdh}\,.
\label{eq:Vexp-new}
\end{align}
In particular, the axions enter at no higher than fourth order.
Maximal supersymmetry uniquely determines the displayed couplings, to the quadratic order in the fermions.  The supersymmetry transformations of the bosonic fields are given by
\begin{align}
\delta_\epsilon{\rm e}&=\bar\epsilon\psi\,,\nonumber\\
\delta_\epsilon\phi&=\bar\epsilon\Gamma_*\lambda\,,\nonumber\\
\delta_\epsilon V_i{}^a&=V_i{}_b\,\bar\epsilon\Gamma^{(a}\chi^{b)}\,,\nonumber\\
\delta_\epsilon a_{ijk}&=-e^\phi V_{[i}{}^aV_j{}^bV_{k]}{}^c
\left(3\bar\epsilon\Gamma_{ab}\chi_c-\bar\epsilon\Gamma_{abc}\Gamma_*\lambda\right),\nonumber\\
\delta_\epsilon A_{ij}&=e^{4\phi}V_{[i}{}^aV_{j]}{}^b
\left(\bar\epsilon\Gamma_{ab}\Gamma_*\psi+4{\rm e}\bar\epsilon\Gamma_{ab}\lambda
-2{\rm e}\bar\epsilon\Gamma_a\Gamma_*\chi_b\right)\nonumber\\
&\quad+e^{3\phi}V_{[i}{}^bV_{j]}{}^c a_{abc}
\left(\bar\epsilon\Gamma^a\Gamma_*\psi+3{\rm e}\bar\epsilon\Gamma^a\lambda
-{\rm e}\bar\epsilon\Gamma_*\chi_a\right).
\label{eq:SUSYbos-new}
\end{align}
The fermionic fields transform as
\begin{align}
\delta_\epsilon\psi&=\mathcal D_t\epsilon+\frac1{24}e^{-\phi}p^{abc}\Gamma_{abc}\epsilon
+{\rm e}\,\mathrm g A\epsilon\,,\nonumber\\
\delta_\epsilon\lambda&=\frac12{\rm e}^{-1}\dot\phi\Gamma_*\epsilon
+\frac1{240}{\rm e}^{-1}e^{-\phi}p^{abc}\Gamma_{abc}\Gamma_*\epsilon
+\mathrm g B\epsilon\,,\nonumber\\
\delta_\epsilon\chi^a&=\frac12{\rm e}^{-1}P^{ab}\Gamma_b\epsilon
+\frac7{80}{\rm e}^{-1}e^{-\phi}p_{bcd}\widetilde\Gamma^{ab,cd}\epsilon
+\mathrm g C^a\epsilon\,.
\label{eq:SUSYfer-new}
\end{align}
The tensors $B$ and $C^a$ describing the fermion shifts of dilatino and fermions are those given in \eqref{eq:YukawaAll-new}, while the gravitino fermion shift is given by the tensor
\begin{equation}
A=-\frac14b\Gamma_*+\frac18b^{abc}\Gamma_{abc}\Gamma_*
-\frac1{32}b^{abcd}\Gamma_{abcd}\Gamma_*\,.
\label{eq:YukA}
\end{equation}
Consequently, the vanishing of these fermion variations directly provides the Killing-spinor system.  

On top of local supersymmetry, the action (\ref{eq:LSUGRA-new}) is invariant under one-dimensional diffeomorphisms, ${\rm SO}(10)$ gauge transformations, acting on indices $i, j, \dots$, and gauge fields $A_{ij}$, as well as local ${\rm SO}(10)$ coset transformations with composite connection $Q^{ab}$ acting on the 'flat' indices $a, b, \dots$, thus in particular on the fermions.

\subsection{BPS backgrounds}

In view of the symmetries of the polarized IKKT model, we are interested in BPS solutions which preserve an
\begin{equation}
{\rm SO}(3)\times{\rm SO}(7)\subset{\rm SO}(10)\,,
\end{equation}
subgroup of the ${\rm SO}(10)$ gauge group. The most general ansatz compatible with this symmetry keeps the einbein, the dilaton, and two scalar fields. All fermions and the gauge fields are set to zero. Explicitly, we parametrize the ${\rm SL}(10)$ matrix $T_{ij}$ as
\begin{equation}
T_{ij}={\rm diag}\bigl(X^{7/10},X^{7/10},X^{7/10},
X^{-3/10},X^{-3/10},X^{-3/10},X^{-3/10},X^{-3/10},X^{-3/10},X^{-3/10}\bigr)\,,
\label{eq:trunc1-new}
\end{equation}
in terms of an ${\rm SO}(3)\times{\rm SO}(7)$ singlet $X$, and the dilaton and singlet axion as
\begin{equation}
e^\phi=e^{\eta/7}X^{-1/20},
\qquad a_{123}=e^{\eta/7}Y\,,
\label{eq:trunc2-new}
\end{equation}
in terms of scalars $\eta$ and $Y$.
The truncated Lagrangian is
\begin{align}
\mathcal L={}&-\frac12{\rm e}^{-1}\frac{\dot X^2+\dot Y^2}{X^2}
-\frac17{\rm e}^{-1}\frac{X\dot X+Y\dot Y}{X^2}\dot\eta
+\frac{10}{49}{\rm e}^{-1}\left(1-\frac1{20}Y^2X^{-2}\right)\dot\eta^2\nonumber\\
&+\frac12\mathrm g^2{\rm e}\,e^{8\eta/7}
\frac{35+42X+3X^2+3Y^2}{X}\,.
\label{eq:LXY-new}
\end{align}
From the ${\rm SO}(10)$ origin of these fields, we can infer the operators dual to these fields from the IKKT BPS multiplet (\ref{eq:B2})
\begin{align}
&X\longleftrightarrow  \mathcal O_{11}+\mathcal O_{22}+\mathcal O_{33}
\;\;\in\;\;{\bf 54}
\,,\nonumber\\
&Y\longleftrightarrow   \mathcal O_{\Omega}
\;\;\in\;\;{\bf 120}
\,,
\label{eq:dualop}
\end{align}
which are singlets under ${\rm SO}(3)\times{\rm SO}(7)$\,. We recall the definition \eqref{eq:Odef} of $\mathcal O_{\Omega}$. 

In this truncation, and with the ansatz (\ref{eq:trunc1-new}), (\ref{eq:trunc2-new}), the Killing spinor equations (\ref{eq:SUSYfer-new})
\begin{equation}
\delta_\epsilon\psi=0\,,
\qquad\delta_\epsilon\lambda=0\,,
\qquad\delta_\epsilon\chi^a=0\,,
\label{eq:KS1}
\end{equation}
reduce to the first order equations
\begin{align}
\frac{dY}{dX}&=-\frac{Y(7X^2+Y^2+8X+9)}{2X\left(2(1-X)(X+3)+Y^2\right)}\,,\nonumber\\
\frac{d\eta}{dX}&=\frac{7\left((X+3)^2+Y^2\right)}{2X\left(2(1-X)(X+3)+Y^2\right)}\,,
\label{eq:KSXY}
\end{align}
together with 
\begin{equation}
\frac{dX}{dt}=\mathrm g\, {\rm e}\,e^{4\eta/7}
\frac{\sqrt X\left(2(X-1)(X+3)-Y^2\right)}{\sqrt{(3+X)^2+Y^2}}\,.
\label{eq:BPSXt}
\end{equation}
These first-order relations are the Euclidean continuation, at $p=-1$, of the spherical D$p$-brane BPS system with ${\rm SO}(2,1)\times{\rm SO}(6-p)$ invariance~\cite{Bobev:2024gqg}. It is straightforward to confirm that they induce all second order field equations of the Lagrangian (\ref{eq:LXY-new}). Throughout the following, we will adopt the gauge ${\rm e}=\frac1{8{\rm g}}$ for the einbein.
For a given solution of (\ref{eq:KSXY}), (\ref{eq:BPSXt}), the Killing spinor solving (\ref{eq:KS1}) is given by
\begin{align}
& \epsilon = f(t)\,\epsilon_0\,,\nonumber\\
&\mathbb P\epsilon_0=0\,,
\qquad
\mathbb P=\frac12(\mathbb I-\Gamma_*)
\left(\mathbb I -
\frac{3+X-\sqrt{(3+X)^2+Y^2}}{Y}\,
\Gpolar\right).
\label{eq:projector-new}
\end{align}
Because $\mathbb P^2=\mathbb P$ and ${\rm tr}\,\mathbb P=16$, all these flows preserve sixteen supercharges.  The precise form of the function $f(t)$ is found from integrating the vanishing gravitino variation $\delta_\epsilon\psi=0$\,. 
With the gauge symmetry broken down to a global ${\rm SO}(3)\times{\rm SO}(7)$ and the supersymmetry given by a spinor in the irreducible ${\bf (2,8)}$ (\ref{eq:1682}), the superisometry group of these backgrounds is given by the simple supergroup $F(4)$.

The BPS equations (\ref{eq:KSXY}), (\ref{eq:BPSXt}) constitute a system of first order differential equations, whose general solution carries two free integration constants (and a trivial third one to be absorbed into a shift of~$t$). It is instructive to expand their solution around the ${\rm SO}(10)$-invariant fixed point $X=1$, $Y=0$,\footnote{Note that this is restrictive. Here we are only interested in BPS solutions that asymptote to this fixed point.}  as
\begin{align}
X(t) ={}& 1 - \initX\,t^{1/4}
+ \frac{7\initX^2}{10}\,t^{1/2}
- \left(\frac{161\initX^3}{400}
+ \frac{\initY^2}{16}\right)t^{3/4}
+ \left(\frac{401\initX^4}{2000}
+ \frac{61\initX\initY^2}{560}\right)t
+ O(t^{5/4})\,,
\nonumber\\[1ex]
Y(t) ={}& \initY\,t^{3/8}
- \frac{11\initX\initY}{20}\,t^{5/8}
+ \frac{61\initX^2\initY}{160}\,t^{7/8}
- \left(\frac{153\initX^3\initY}{640}
+ \frac{3\initY^3}{224}\right)t^{9/8}
+ O(t^{11/8})\,,
\nonumber\\[1ex]
e^{\eta(t)/7} ={}& t^{-1/4}
- \frac{\initX}{20}
- \frac{\initX^2}{160}\,t^{1/4}
+ \left(\frac{21\initX^3}{3200}
- \frac{3\initY^2}{448}\right)t^{1/2}
+ O(t^{3/4})\,,
\label{eq:XYetaExp}
\end{align}
with free integration constants $\initX,\initY\geq 0$.

\subsubsection{D($-1$) instanton solution and pure vev-deformation}

The unique fully ${\rm SO}(10)$-symmetric configuration is recovered at
\begin{align}
\initX=\initY=0\quad
\Longrightarrow\quad
& \mathring{X}=1\,,\; \mathring{Y}=0\,,\; e^{\mathring{\eta}/7}=t^{-1/4}\,,
\nonumber\\
& \mathring{\epsilon} = t^{5/16}\,\epsilon_0\,,\quad
\Gamma_*\,\epsilon_0 = \epsilon_0
\,.
\label{eq:solSO10}
\end{align}
This is the dual background to the undeformed IKKT model. Its IIB uplift gives the near-horizon solution for coincident D$(-1)$ \cite{Gibbons:1995vg} instantons which sit at $t\rightarrow \infty$. The UV boundary is at $t=0$. Note that here and in the following a circle superscript denotes background values for the fields.

Other distinguished solutions to the BPS equations (\ref{eq:KSXY}), (\ref{eq:BPSXt}) are the cases of either $\initX$ or $\initY$ vanishing. For $\initY=0$, the solution can be given in analytical form
\begin{align}
\mathring{X} &= \frac{\tilde t-1}{\tilde t}\,,\quad
\mathring{Y} = 0\,,
\quad
e^{\mathring{\eta}/7}
= m^{2/3}\,\tilde t^{1/4}(\tilde t-1)^{3/4}\,,\quad
\frac{d\tilde t}{dt}
= -\frac{m^{8/3}}{4}\,\tilde t^{3/2}(\tilde t-1)^{7/2}\,,\nonumber\\[1ex]
\mathring{\epsilon}
&= \tilde t^{-3/8}(\tilde t-1)^{-7/8}\epsilon_0\,,
\qquad \Gamma_*\,\epsilon_0=\epsilon_0\,,\label{eq:vevdefsol}
\end{align}
with coordinate $\tilde t>1$ and scale parameter $m$, related to the expansion (\ref{eq:XYetaExp}) by $\initX=m^{2/3}$. This is an ${\rm SO}(3)\times{\rm SO}(7)$\ invariant solution in the class of BPS solutions found in~\cite{Ciceri:2025maa}, with an asymptotic UV behavior at $\tilde t\rightarrow \infty$ that recovers the SO(10)-symmetric solution \eqref{eq:solSO10}. Its IIB uplift yields a flat Euclidean metric with a non-trivial axio-dilaton describing an ${\rm SO}(3)\times{\rm SO}(7)$-invariant continuous distribution of instantons. We expect it to be dual to a pure vev-deformation of the IKKT model associated with an operator transforming in the ${\bf 54}$. In particular, the scalar $Y$ is not excited in this background. The flow runs over $0<\mathring X<1$ and ends at the fixed point $(X,Y)=(0,0)$ as $\tilde t\to1$, \textit{i.e.}\ $t\to\infty$, see figure~\ref{fig:flows}.

\subsubsection{Polarized solution}

On the other hand, setting $\initX=0$ in the expansion \eqref{eq:XYetaExp} yields a unique solution driven at leading order by the scalar $Y$, dual to the operator $\mathcal O_{\Omega}$ in the ${\bf 120}$ of $\mathrm{SO}(10)$. Its backreaction induces a nontrivial excitation of $X$ at subleading orders. This solution may be written in the explicit form
\begin{align}
\mathring{X}
= \,&{} \frac{(1-\tau)^2}{(1+\tau)^2}\,,\;\,
\mathring{Y}
= \frac{8\tau^{1/2}(1+\tau+\tau^2)}
{(1+\tau)^2(1-\tau)}\,,\;\,
e^{\mathring{\eta}/7}= \Big(\frac{\mu}{16\,{\rm g}}\Big)^{2/3}\frac{1-\tau}
{\tau^{1/3}(1+\tau^2)^{1/6}}\,,
\nonumber\\[1ex]
\frac{d\tau}{dt}
=\,&{} \frac34\Big(\frac{\mu}{16\,{\rm g}}\Big)^{8/3}\frac{(1-\tau)^3(1+\tau^2)^{1/3}}
{\tau^{1/3}(1+\tau)}\,,
\label{eq:BPSpol}
\end{align}
with coordinate $\tau$ and scale parameter $\mu>0$, related to the expansion \eqref{eq:XYetaExp} by $\initX=0$, $\initY=\mu/(2{\rm g})$, and the Killing spinor given by
\begin{equation}
\mathring{\epsilon}
= \frac{\tau^{5/12}(1+\tau^2)^{1/3}}
{(1+\tau)^{1/2}(1-\tau)^{3/2}}
\left(1-\tau^{1/2}\Gpolar\right)\epsilon_0\,,
\qquad \Gamma_*\epsilon_0=\epsilon_0\,.
\label{eq:BPSpolKS}
\end{equation}
This polarized solution lies on the curve $\mathring W=0$ in the $(X,Y)$ plane, where \cite{Bobev:2024gqg}
\begin{equation}
W:=(1-X)(3+X)^2-XY^2\,.
\label{eq:Wcurve}
\end{equation}
It approaches the ${\rm SO}(10)$ UV fixed point as $\mathring Y^2\simeq16(1-\mathring X)$ and reaches $\mathring X=0$, $\mathring Y\to\infty$ as $\tau\to1$, \textit{i.e.}\ $t\to\infty$.

This is the solution we will be mostly focusing on in this paper. According to its dominant behavior in the expansion \eqref{eq:XYetaExp}, we expect it to be dual to an ${\rm SO}(3)\times{\rm SO}(7)$ invariant operator deformation of the IKKT model by the operator $\mathcal O_{\Omega}$ from the ${\bf 120}$. This is precisely the signature of the polarized IKKT model \eqref{eq:IKKTpolar}.

\subsubsection{General solutions}

For generic values of both integration constants, the solutions combine the polarization driven by $\initY$ with the vev-deformation driven by $\initX$. The BPS equations \eqref{eq:KSXY}, \eqref{eq:BPSXt} are invariant under the rescaling $t\to\lambda^{-4}t$, $e^{\eta/7}\to\lambda\,e^{\eta/7}$, which maps the solution with integration constants $(\initX,\initY)$ to the one with $(\lambda\initX,\lambda^{3/2}\initY)$.\footnote{From the IIB supergravity point of view, this is a combination of a trombone scaling and an SL(2) scaling.} The solutions are therefore labelled, up to this rescaling, by
\begin{equation}
k:=\frac{\initY^2}{\initX^3}\geq0\,,
\label{eq:k}
\end{equation}
with the pure vev-deformation and the polarized solution as the limits $k=0$ and $k\to\infty$, respectively.\footnote{
This general solution has been given in analytic form recently in \cite{Gibson:2026mqk} with its integration constant $K$ related to (\ref{eq:k}) by $k=\frac{4}{8-K}$\,.
} 
They approach the ${\rm SO}(10)$ point as $\mathring Y^2\simeq k(1-\mathring X)^3$.\footnote{For large $k$ the behavior $\mathring Y^2\simeq k(1-\mathring X)^3$ only holds for $1-\mathring X\lesssim 4/\sqrt k$, beyond which the solutions follow the polarized one, $\mathring Y^2\simeq16(1-\mathring X)$.} Along each of them $e^{2\mathring\eta/7}\mathring Y^2/\mathring W=\initY^2/(16\initX)$ is constant, with $W$ as in \eqref{eq:Wcurve}. This constant diverges for the polarized solution, on which $\mathring W=0$.

\begin{figure}[!htbp]
\centering
\includegraphics[width=0.6\textwidth]{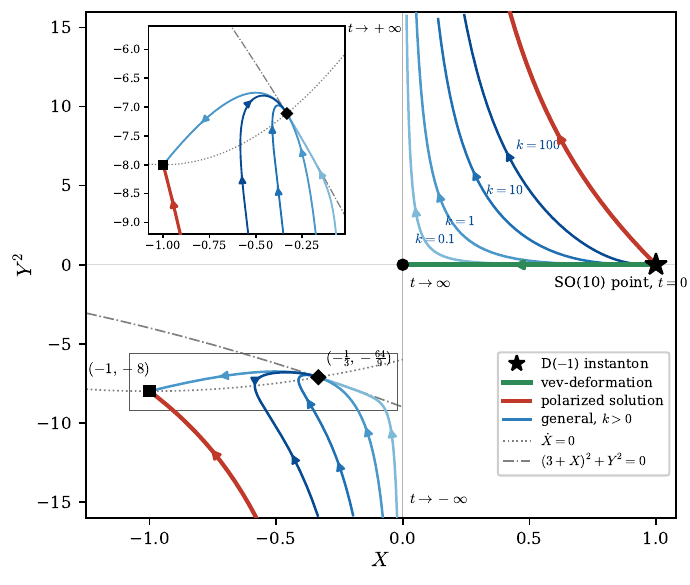}
\captionsetup{width=0.95\textwidth,font=footnotesize}
\caption{1/2-BPS solutions in the $(X,Y^2)$ plane, with arrows pointing towards increasing $t$. All start at $t=0$ from the ${\rm SO}(10)$ fixed point (star), \textit{i.e.}\ the D($-1$) instanton \eqref{eq:solSO10}. The pure vev-deformation (green) ends at the $(X,Y)=(0,0)$ fixed point (dot). The polarized solution (red) and the general solutions with $0<k<\infty$ (blue) cross $\mathring X=0$ at $t=\pm\infty$, with $\mathring Y$ imaginary for $\mathring X<0$. The former ends at the $(X,Y^2)=(-1,-8)$ fixed point (square), the latter at $(X,Y^2)=(-\frac13,-\frac{64}{9})$ (diamond), except for $k=1$, which continues to $(-1,-8)$. The inset magnifies the region around these end points.}
\label{fig:flows}
\end{figure}

For $0<k<\infty$ the solutions flow through $0<\mathring X<1$, between the vev-deformation and the polarized solution, and reach $\mathring X=0$, $\mathring Y\to\infty$ as $t\to\infty$. For $k=1$ the solution takes the closed form
\begin{equation}
\mathring Y^2=\frac{(1-\mathring X)^3}{\mathring X}\,,\qquad
e^{\mathring\eta/7}=\frac{\initX}{\sqrt2}\,\frac{\sqrt{\mathring X(1+\mathring X)}}{1-\mathring X}\,.
\end{equation}
The polarized solution and the solutions with $0<k<\infty$ can be continued through $\mathring X=0$, which is reached at $t=\pm\infty$, to $\mathring X<0$ with imaginary $\mathring Y$ and $e^{\mathring\eta/7}$, and real $t$.\footnote{ For the polarized solution, this branch is given by \eqref{eq:BPSpol} with $\tau=e^{i\varphi}$, $0<\varphi<\pi/2$, on which $\mathring X=-\tan^2(\varphi/2)$.} The former then ends at the fixed point $(X,Y^2)=(-1,-8)$, the latter at $(X,Y^2)=(-\frac13,-\frac{64}{9})$, except for $k=1$, which passes through this point and also ends at $(-1,-8)$. The various solutions are displayed in figure~\ref{fig:flows}.

\subsection{Uplift of BPS backgrounds}

For the half-BPS solutions described by \eqref{eq:KSXY}, \eqref{eq:BPSXt}, the uplift to the fields of IIB supergravity in Einstein frame takes the form \cite{Ciceri:2026hxr}
\begin{align}
ds_{10}^2&=\frac{e^{\eta/7}}{\mathrm g^2\mathcal P^{1/4}\mathcal Q^{3/4}}
\left(\mathrm g^2{\rm e}^2e^{8\eta/7}dt^2+d\theta^2
+\mathcal P\cos^2\theta\,d\Omega_2^2
+\mathcal Q\sin^2\theta\,d\Omega_6^2\right),\nonumber\\
e^\Phi&=e^{-4\eta/7}\sqrt{\mathcal P\mathcal Q}\,,\nonumber\\
\mathcal X&=-\frac{e^{4\eta/7}X^{1/2}}{\sqrt{(3+X)^2+Y^2}}
\left(\frac1{\mathcal P}+\frac3{\mathcal Q}
+\frac{Y^2}{2X^2}\left(\cos^2\theta-\frac14\right)\right),\nonumber\\
B_{(2)}&=-e^{-\eta/7}\frac{Y\mathcal P}{\mathrm g^2X}\cos^3\theta\,\mathrm{vol}_2\,,\nonumber\\
C_{(2)}&=\frac{e^{3\eta/7}Y\mathcal P\cos^3\theta}
{\mathrm g^2X^{1/2}\sqrt{(3+X)^2+Y^2}}
\left(\frac5{2\mathcal Q}+\frac1{6X\mathcal P}-\frac{Y^2}{8X^2}\right)\mathrm{vol}_2\,.
\label{eq:UA-new}
\end{align}
This map is specific to the half-BPS sector and may require modification for less-supersymmetric solutions of \eqref{eq:LXY-new}. Here $0\leq\theta\leq\pi/2$ and
\begin{align}
\mathcal P&=\frac{X}{X\sin^2\theta+(X^2+Y^2)\cos^2\theta},
\qquad
\mathcal Q=\frac{X}{\sin^2\theta+X\cos^2\theta}.
\end{align}
The notation $d\Omega_n^2$ and $\mathrm{vol}_n$ refers to the unit $S^n$ metric and volume form.  The RR four-form is vanishing. Using the one-dimensional BPS relations, these fields satisfy the Euler--Lagrange equations of Euclidean type-IIB supergravity derived from
\begin{equation}
\mathcal L_{\rm IIB}=\frac1{2\kappa_{10}^2}\int d^{10}x
\left(R-\frac12(\partial\Phi)^2+\frac12e^{2\Phi}(\partial\mathcal X)^2
-\frac1{12}e^{-\Phi}(H_{(3)})^2+\frac1{12}e^\Phi(F_{(3)})^2\right).
\end{equation}
The NS--NS and RR three-form strengths are, respectively,
\begin{align}
H_{(3)}=dB_{(2)}\,,
\qquad
F_{(3)}=dC_{(2)}-\mathcal X H_{(3)}\,.
\label{eq:fortyone-new}
\end{align}

\subsubsection{Electrostatic potential formulation}
For $\initY>0$, these ten-dimensional half-BPS solutions can also be conveniently described in terms of a potential. In the coordinates
\begin{equation}
z=\frac{\mu\,e^{-\eta/7}\,(3+X)}{8\,{\rm g}^2\,Y}\,\cos\theta\,,\qquad
\rho=\frac{\mu\,e^{-\eta/7}\sqrt{X}\sqrt{(3+X)^2+Y^2}}{8\,{\rm g}^2\,Y}\,\sin\theta\,,
\label{eq:pol-zrho}
\end{equation}
with $\mu=2{\rm g}\initY$ as above, the one-dimensional BPS relations \eqref{eq:KSXY} imply that
\begin{equation}
V=\frac{\mu^4\cos\theta\,e^{3\eta/7}\,Y\big(2(X+1)\cos2\theta+X-1\big)}{64\,{\rm g}^2\,X^{3/2}\sqrt{(3+X)^2+Y^2}}
\label{eq:pol-Vgen}
\end{equation}
satisfies the Laplace equation of a four-dimensional axially symmetric system,\footnote{The remaining two directions are not part of the ten-dimensional spacetime.}
\begin{equation}
\partial_{z}^2V+\partial_\rho^2 V+\frac{2}{\rho}\partial_\rho V=0\,.
\end{equation}
As explained in detail in \cite{Komatsu:2024bop,Komatsu:2024ydh,Najafian:2026wld}, $V(z,\rho)$ can be interpreted as the electrostatic potential associated to a distribution of conducting balls centered along the vertical axis $z$. 

\subsubsection{Uplift of the polarized solution}

The uplift of the pure vev-deformation \eqref{eq:vevdefsol} has been discussed in \cite{Ciceri:2025maa}. Here we only focus on the polarized solution \eqref{eq:BPSpol}. For this solution $\sqrt{\mathring X\big((3+\mathring X)^2+\mathring Y^2\big)}=3+\mathring X$, so that \eqref{eq:pol-zrho} is a polar map, $z=r\cos\theta$, $\rho=r\sin\theta$, with
\begin{equation}
\frac{r^6}{r_0^6}=\frac{1+\tau^2}{2\tau}=\frac{1+\mathring X}{1-\mathring X}\,,\qquad
r_0^6=\frac{\mu^2}{128\,{\rm g}^8}\,.
\end{equation}
Inverting, the one-dimensional fields depend only on $r^2=z^2+\rho^2$,
\begin{equation}
\mathring X=\frac{r^6-r_0^6}{r^6+r_0^6}\,,\qquad
\mathring Y=\frac{2\sqrt2\,r_0^3\,(2r^6+r_0^6)}{(r^6+r_0^6)\sqrt{r^6-r_0^6}}\,,\qquad
e^{\mathring\eta/7}=\frac{{\rm g}^2\sqrt{r^6-r_0^6}}{r}\,.
\label{eq:pol-dict}
\end{equation}
The coordinates cover the quarter plane $z\geq0$, $\rho\geq0$ with the origin removed. The UV fixed point $(X,Y)=(1,0)$ lies at $r\to\infty$, the locus $\mathring X=0$ on the quarter circle $r=r_0$, and the IR fixed point $(X,Y)=(-1,\pm2\sqrt2\,i)$ at the origin. The exterior $r>r_0$ thus corresponds to $0<\mathring X<1$, and the interior $0<r<r_0$ to the continuation $-1<\mathring X<0$.

In the gauge ${\rm e}=1/(8{\rm g})$, the flow equation in \eqref{eq:BPSpol} gives ${\rm g}\,{\rm e}\,e^{4\mathring\eta/7}dt=-dr/r$, so that the base in \eqref{eq:UA-new} becomes conformally flat. With
\begin{equation}
\begin{aligned}
&\Delta_P=r^2\big(r^6-r_0^6\big)^2+6\,r_0^6\,z^2\big(5r^6+r_0^6\big)\,,\qquad
\Delta_Q=r^8-r_0^6\big(z^2-\rho^2\big)\,,\\[1ex]
&\mathring{\mathcal P}=\frac{r^2\big(r^6-r_0^6\big)^2}{\Delta_P}\,,\qquad
\mathring{\mathcal Q}=\frac{r^2\big(r^6-r_0^6\big)}{\Delta_Q}\,,
\end{aligned}
\end{equation}
the uplift \eqref{eq:UA-new} of the polarized solution then reads
\begin{align}
d\mathring s_{10}^2&=\frac{\sqrt{r^6-r_0^6}}{r^3\,\mathring{\mathcal P}^{1/4}\mathring{\mathcal Q}^{3/4}}
\Big(dz^2+d\rho^2+\mathring{\mathcal P}\,z^2\,d\Omega_2^2+\mathring{\mathcal Q}\,\rho^2\,d\Omega_6^2\Big)\,,\nonumber\\[1ex]
e^{\mathring\Phi}&=\frac{r^4\sqrt{\mathring{\mathcal P}\mathring{\mathcal Q}}}{{\rm g}^8\big(r^6-r_0^6\big)^2}\,,\qquad
\mathring{\mathcal X}=-\frac{{\rm g}^8}{2r^6}\Big[2r^{14}+\big(20z^2-6r^2\big)r_0^6\,r^6+\big(r^2-8z^2\big)r_0^{12}\Big]\,,\nonumber\\[1ex]
\mathring B_{(2)}&=-\frac{\mu\,\big(2r^6+r_0^6\big)\,z^3}{4\,{\rm g}^8\,\Delta_P}\;\mathrm{vol}_2\,,\nonumber\\[1ex]
\mathring C_{(2)}&=\frac{\mu\,z^3}{24\,r^6\,\Delta_P}
\Big[r^2\big(8r^{18}-20r_0^6r^{12}-20r_0^{12}r^6+5r_0^{18}\big)
+12\,r_0^{12}\,z^2\big(4r^6-r_0^6\big)\Big]\;\mathrm{vol}_2\,,
\label{eq:pol-10d}
\end{align}
and the electrostatic potential \eqref{eq:pol-Vgen} becomes
\begin{equation}
\mathring V=\frac{\mu^5}{128}\,z\big(z^2-\rho^2\big)-\frac{\mu^7}{2^{15}\,{\rm g}^8}\,\frac{z}{r^4}\,.
\label{eq:pol-V}
\end{equation}
It consists only of the cubic background term, fixed by $\mu$, and of a point dipole at the origin, with dipole moment proportional to $\mu^7{\rm g}^{-8}$. In the multipole expansion of $V$ discussed in \cite{Komatsu:2024bop}, all higher multipole moments thus vanish. In the picture of conducting balls, the polarized solution is the degenerate configuration without balls of finite size.

The background term dominates as $r\to\infty$, where the solution approaches the near-core D-instanton in flat $\mathbb R^3\times\mathbb R^7$, with $z$ and $\rho$ the radial coordinates of the two factors, deformed by the polarization
\begin{equation}
d\mathring C_{(2)}=\mu\,{\rm vol}(\mathbb R^3)+{\cal O}(r^{-4})\,,\qquad
e^{-\mathring\Phi}+\mathring{\mathcal X}=\frac{\mu^2}{64}\,\big(3z^2+\rho^2\big)+{\cal O}(r^{-4})\,.
\label{eq:pol-UV}
\end{equation}
The quadratic term has the same $3:1$ anisotropy as the mass terms in \eqref{eq:IKKTpolar}. All corrections to the D-instanton background are controlled by $(r_0/r)^6=\mu^2/(128{\rm g}^8r^6)$.

When moving inwards, the $S^2$ and $S^6$ shrink and the dilaton blows up on the quarter circle $r=r_0$. In the interior, \eqref{eq:pol-dict}--\eqref{eq:pol-V} hold with $\sqrt{r^6-r_0^6}=-i\sqrt{r_0^6-r^6}$. This means the metric and, in part of the domain, the dilaton are then complex, while the axion, the two-forms and $\mathring V$ remain real. We take the singular locus and the complex interior as a sign that this degenerate configuration is not a regular solution on its own. Its asymptotic region, on the other hand, carries the characteristic features of the polarization, see \eqref{eq:pol-UV}. As we will see in the following, this is sufficient for our purpose of relating the asymptotic data of the bulk fluctuations to the BPS operators of the polarized matrix model, since we will keep all integration constants of the fluctuations and impose no condition in the interior.

\section {Asymptotic expansion and fluctuation equations}\label{sec:asymptotic}


In this section, we compute and solve the linearized fluctuation equations around the undeformed and the polarized BPS backgrounds.
As a warm-up, we compute the linearized, bosonic and fermionic, field equations around the ${\rm SO}(10)$ invariant background. We give their exact solutions in terms of a number of integration constants which encode the physical degrees of freedom of this sector. We then proceed to the fluctuation equations around the polarized ${\rm SO}(3)\times{\rm SO}(7)$ invariant background. Again, we give full analytic solutions to these equations together with their asymptotic expansion at the boundary, in terms of their independent integration constants.

For the supersymmetry analysis, we will also expand the supersymmetry parameter as
\begin{equation}
\epsilon(t) = \mathring{\epsilon}(t) + \eta(t)\,,
\label{eq:susyfluc}
\end{equation}
where $\mathring{\epsilon}(t)$ is the Killing spinor of the respective background, and $\eta(t)$ is a free local parameter of the order of the fluctuations. The fluctuations will then transform under the {\em global} supersymmetry induced by the constant spinor $\epsilon_0$ parametrizing the Killing spinor, as well as under {\em local} shift symmetries by $\eta(t)$\,. We use the latter symmetry to eliminate some of the unphysical modes within the gravitino/dilatino sector. The structure of supermultiplets w.r.t.\ the global supersymmetry $\epsilon_0$ will then be discussed in section~\ref{sec:supermultiplets} below.

\subsection{${\rm SO}(10)$ BPS domain wall}

The ${\rm SO}(10)$ invariant (unpolarized) supersymmetric solution \eqref{eq:solSO10} yields the explicit background fields
\begin{equation}
  \mathring{\rm e}=\frac1{8{\rm g}},\quad 
  \mathring{\cT}_{ab}=\delta_{ab}\,,\quad  \mathring{a}_{abc}=0\,,\quad
 e^{ \mathring\phi}=t^{-1/4}\,,\quad
  \mathring{A}_{ab}=0\,,
  \label{eq:so10background}
\end{equation}
On this background, the non-vanishing Yukawa tensors are the ${\rm SO}(10)$ singlets
\begin{equation}
 \mathring{A}=-\frac{5}{2t}\,\Gamma_*\,,\quad
 \mathring{B}=\frac1t\,,\quad
  \mathring{E}=-\frac{210}{t}\,\Gamma_*\,, \quad
 \mathring{E}^{ab}=\frac1t\,\delta^{ab}\,\Gamma_*\,.
 \label{eq:YukawaSO10}
\end{equation}
The UV boundary lies at $t=0$\,.

\subsubsection*{Bosonic fluctuations}

Let us first consider the bosonic fluctuations around this background.
We parametrize these fluctuations by
\begin{align}
 V_i{}^a = \,&\delta_i{}^a + s_i{}^a  \,,\quad s_{ab}=s_{ba}\,,\quad s_a{}^a=0\,,\nonumber\\
 \phi=\,& \mathring\phi+v\,,\quad
 {\rm e}=\mathring{\rm e}\,(1+n)\,,\quad 
  a_{abc}=\alpha_{abc}\,,\quad
A_{ab}=\beta_{ab}\,.
 \label{eq:so10fluctansatz}
\end{align}
They appear in ${\rm SO}(10)$ representations
\begin{equation}
 s_{ab}\in\mathbf{54}\,,\quad
 \alpha_{abc}\in\mathbf{120}\,,\quad
 \beta_{ab}\in\mathbf{45}\,,\quad
 v, n \in \mathbf{1}\,.
\end{equation}
The bosonic Lagrangian expanded to quadratic order in these fluctuations takes the form
\begin{align}
g^{-1}\,{\cal L}^{(2)}_{{\rm bos}}
 &=-8\,\dot s_{ab}\dot s_{ab}
 +\frac{3}{2}\,t^{-2}\,s_{ab}s_{ab}-\frac2{3} \,t^{1/2}
   \dot\alpha_{abc}\dot\alpha_{abc}
 +\frac{1}{32}\,t^{-3/2}\,\alpha_{abc}\alpha_{abc}\,,\nonumber\\
 &{} + 80 \, \left(\dot v^2+\frac{n\dot v}{2t}
 +\frac{n^2}{16t^2}\right)
 +\frac{5}{t^2}\left(32v^2+8nv\right)\,.
\label{eq:Lfluc-bos}
\end{align}
Around this background, there is no quadratic term for the gauge field fluctuations $\beta_{ab}$, which we may set to zero by linearized ${\rm SO}(10)$ gauge transformations.

For the non-singlet fields, the fluctuation equations can be solved explicitly as
\begin{align}
s_{ab}(t)=\,& b^{(2)}_{ab}\,t^{1/4}+b^{(6)}_{ab}\,t^{3/4}\,,\nonumber\\
\alpha_{abc}(t)=\,&a^{(1)}_{abc}\,t^{1/8} +a^{(3)}_{abc}\,t^{3/8}\,,
 \label{eq:so10series}
\end{align}
with the integration constants $\big\{b^{(2)}, b^{(6)}, a^{(1)}, a^{(3)}\big\}$.
In the singlet sector,
the lapse constraint is
\begin{equation}
 n =-4t\dot v-4v\,,
 \label{eq:so10singconstraint}
\end{equation}
and implies the second order equation for $v$. A radial diffeomorphism with
parameter $\xi(t)$ acts as
\begin{equation}
 \delta_\xi v=-\frac{\xi}{4t}\,,\qquad
 \delta_\xi n=\dot\xi\,.
\end{equation}
The choice $\xi=4tv$ therefore sets $v=n=0$, the singlet sector has no
physical integration constant.

\subsubsection*{Fermionic fluctuations}

Around the ${\rm SO}(10)$ background, the fermionic Lagrangian to quadratic order reduces to
\begin{align}
{\cal L}^{(2)}_{{\rm ferm}}={}&20\bar\lambda\dot\lambda+2\bar\chi^a\dot\chi_a
+\frac{40{\rm g}}{t}\bar\psi\Gamma_*\lambda
-\frac{40{\rm g}}{t}\bar\lambda\psi
-\frac{105}{4t}\bar\lambda\Gamma_*\lambda
+\frac{1}{8t}\bar\chi_a\Gamma_*\chi_a\,,
\label{eq:Lfluc-ferm}
\end{align}
with the Yukawa tensors  \eqref{eq:YukawaSO10}. For the physical fermions in the ${\bf 144}$, the field equations read
\begin{equation}
 \dot\chi_a+\frac1{16t}\Gamma_*\chi_a=0\,.
\end{equation}
and are solved by
\begin{equation}
 \chi_a(t)=t^{-1/16}\kappa_{a+}+t^{1/16}\kappa_{a-}\,,\qquad
 \Gamma_*\kappa_{a\pm}=\pm\kappa_{a\pm}\,,\quad
 \Gamma^a\kappa_{a\pm}=0\,,
  \label{eq:so10chiseries}
\end{equation}
where in general $\pm$ subscripts are used to denote the chirality of spinors.

The gravitino $\psi$ appears as a Lagrange multiplier in \eqref{eq:Lfluc-ferm} and yields the constraint 
\begin{equation}
 (\mathbb I+\Gamma_*)\lambda=0 \quad\Longrightarrow\quad
 \lambda_+=0\,.
\end{equation}
In turn, the dilatino equation is
\begin{equation}
 40\dot\lambda-\frac{105}{2t}\Gamma_*\lambda
 -\frac{40{\rm g}}{t}(\mathbb I-\Gamma_*)\psi=0\quad
 \Longrightarrow\quad
 40\dot\lambda_-+\frac{105}{2t}\lambda_-
 -\frac{80{\rm g}}{t}\,\psi_-=0
 \,.
 \label{eq:dilatino10}
\end{equation}
Under local supersymmetry \eqref{eq:susyfluc} with parameter $\eta(t)$, the fields in this sector transform as
\begin{equation}
 \delta_\eta\lambda
 =\frac{{\rm g}}{t}(1-\Gamma_*)\,\eta
 \qquad
 \delta_\eta\psi=\dot\eta-\frac{5}{16t}\Gamma_*\eta \,,
\end{equation}
which can be used to set to zero the remaining part of the dilatino: $\lambda_-=0$.  
The remaining supersymmetry parameter $\eta_+$ can then be used to remove the
remaining part $\psi_+$ of the gravitino.
There are no independent fermionic constants in this sector.
A general variation of the fields has to be accompanied by a compensating supersymmetry transformation in order to preserve the gauge $\lambda_-=0$. Because the background has $\mathring{P}^{ab}=\mathring{p}_{abc}=\mathring{C}^a=0$, this
compensator does not alter the linear transformation of $\chi_a$.  Similarly, a compensating ${\rm SO}(10)$ transformation restores
$\beta_{ab}=0$ without changing the transformation of the other fields at linear order.

\subsection{${\rm SO}(3)\times {\rm SO}(7)$ BPS domain wall}

We now perform a similar analysis and asymptotic expansion for the fluctuations around the polarized BPS background \eqref{eq:BPSpol}. While it preserves the same amount of supersymmetry, as the $\mathrm{SO}(10)$ invariant domain wall, it breaks the remaining bosonic symmetries down to
\begin{equation}
{\rm SO}(10) \longrightarrow {\rm SO}(3) \times {\rm SO}(7)
\,.
\label{eq:10-37}
\end{equation}
The supergravity fluctuations break according to the pattern \eqref{eq:branchB2}.
As for the polarized IKKT model, we use indices
\begin{equation}
\indI{1},\indI{2},\indI{3}=1,2,3\,,\qquad
\indA{1},\indA{2},\indA{3}=4,\ldots,10\,,
\end{equation}
in order to label the ${\rm SO}(3)$ and ${\rm SO}(7)$ vector indices, respectively.
In the $\mathrm{SO}(10)$ analysis of the previous section, residual diffeomorphisms and local supersymmetry acted only on nonphysical fluctuations, which could then be removed without affecting the physical modes. After the breaking \eqref{eq:10-37}, singlet and fundamental spinor representations also occur among the physical fluctuations and transform under these local symmetries. Identifying the physical modes therefore requires disentangling them from the gauge fluctuations and the associated constraints. 

In the constant-einbein gauge, and with the shorthand $\hat\mu:=\mu/(16{\rm g})$, the polarized background \eqref{eq:BPSpol} takes the form
\begin{align}
 \mathring{\mathrm e}=\,& \frac{1}{8\mathrm g}\,,\qquad
 e^{\mathring\phi}=
 \frac{\hat\mu^{2/3}(1+\tau)^{1/10}(1-\tau)^{9/10}}
 {\tau^{1/3}(1+\tau^2)^{1/6}}\,,
 \nonumber\\[1ex]
 \mathring{\mathcal T}_{\indI{1}\indI{2}}=\,& \frac{(1-\tau)^{7/5}}{(1+\tau)^{7/5}}\,\delta_{\indI{1}\indI{2}}\,,\quad
 \mathring{\mathcal T}_{\indA{1}\indA{2}}= \frac{(1-\tau)^{-3/5}}{(1+\tau)^{-3/5}}\,\delta_{\indA{1}\indA{2}}\,,\quad
\mathring a_{123}^{\mathrm{curved}}=
 \frac{8\hat\mu^{2/3}\,\tau^{1/6}(1+\tau+\tau^2)}
 {(1+\tau)^2(1+\tau^2)^{1/6}}
 \,,
\label{eq:so37-background}
\end{align}
with $\tau$ defined in \eqref{eq:BPSpol} as a function of $t$. We choose its integration constant such that the UV boundary lies at $t=0$, where $\tau$ behaves as
\begin{equation}
\tau(t) = \hat\mu^2\,t^{3/4} - \frac{12}{7}\,\hat\mu^4\,t^{3/2} + {\cal O}(\hat\mu^6t^{9/4})\,.
\label{eq:rhomt}
\end{equation}
Near the UV boundary, the polarized solution \eqref{eq:so37-background} approaches the undeformed ${\rm SO}(10)$ invariant domain wall solution \eqref{eq:so10background}.
All gauge fields vanish in the background. 
The axion component displayed in \eqref{eq:so37-background} refers to the component in curved
$\mathrm{SO}(10)$ indices $a_{ijk}$, to be distinguished from its flattened version $a_{abc}$ from \eqref{eq:axionSO-new}.
In the following, we rescale the Killing spinor \eqref{eq:BPSpolKS} by $\hat\mu^{-5/6}$, so that near the UV boundary it takes the form
\begin{equation}
\mathring\epsilon=t^{5/16}\Big(1-\hat\mu\,t^{3/8}\,\Gpolar+{\cal O}\big(\hat\mu^2t^{3/4}\big)\Big)\,\epsilon_0\,,
\label{eq:mkillingspinorexact}
\end{equation}
with the same constant spinor $\epsilon_0$ as for the ${\rm SO}(10)$ background \eqref{eq:solSO10}.


\subsubsection*{Bosonic fluctuations}

As before, we write $V=\mathring V\exp(s)$, where $s$ is symmetric and traceless, and
parametrize the other bosonic fluctuations as
\begin{equation}
 \phi= \mathring\phi+v\,,\quad
 {\rm e}=\mathring{\rm e}\,(1+n)\,,\quad 
  a_{ijk}=\mathring{a}_{ijk} + \alpha_{ijk}\,,\quad
A_{ij}=\beta_{ij}\,.
 \label{eq:so37fluctansatz}
\end{equation}
We further decompose the fluctuations according to their breaking \eqref{eq:branchB2}
into ${\rm SO}(3)$ and ${\rm SO}(7)$ representations
\begin{align}
s_{ab} \,:\quad & s_{\indI{1}\indI{2}}=-\frac13u\,\delta_{\indI{1}\indI{2}}
+\sigma_{\indI{1}\indI{2}}\,,\quad
s_{\indA{1}\indA{2}}=\frac17u\,\delta_{\indA{1}\indA{2}}
+\sigma_{\indA{1}\indA{2}}\,,\quad
s_{\indI{1}\indA{1}}\,,\nonumber\\
\alpha_{ijk}\,:\quad & \alpha_{123}=a\,,\quad
\alpha_{\indI{1}\indI{2}\indA{1}}=
\epsilon_{\indI{1}\indI{2}\indI{3}}
z_{\indI{3}\indA{1}}\,,\quad
\alpha_{\indI{1}\indA{1}\indA{2}}\,,\quad
\alpha_{\indA{1}\indA{2}\indA{3}}\,,
\nonumber\\
\beta_{ij}\,:\quad& 
\beta_{\indI{1}\indI{2}}\,,\quad
\beta_{\indA{1}\indA{2}}\,,\quad
\beta_{\indI{1}\indA{1}}\,,
\label{eq:so37-component-embedding}
\end{align}
where  $\sigma_{\indI{1}\indI{2}}$ and
$\sigma_{\indA{1}\indA{2}}$ are separately traceless, and all axion
components carry curved indices.
Thus $u=s_{\indA{1}}{}^{\indA{1}}=-s_{\indI{1}}{}^{\indI{1}}$.
The unindexed $b^{(2)}$ and $b^{(6)}$ below are its integration constants.

Let us first consider the fluctuations $\{\sigma_{\indI{1}\indI{2}}, \sigma_{\indA{1}\indA{2}},
\alpha_{\indI{1}\indA{1}\indA{2}}, \alpha_{\indA{1}\indA{2}\indA{3}}\}$  whose representations appear only once in the spectrum, such that their fluctuation equations are automatically diagonal.
Expanding the Lagrangian to quadratic order in these fluctuations leads to
\begin{align}
\mathrm g^{-1}\mathcal L^{(2)}_{{\rm bos}}={}&
-8\bigl(\dot\sigma_{\indI{1}\indI{2}}\dot\sigma_{\indI{1}\indI{2}}
 +\dot\sigma_{\indA{1}\indA{2}}\dot\sigma_{\indA{1}\indA{2}}\bigr)
-\frac{2\tau^{2/3}\,(1+\tau^2)^{1/3}}{\hat\mu^{4/3}}
 \left(\frac{\dot\alpha_{\indI{1}\indA{1}\indA{2}}
 \dot\alpha_{\indI{1}\indA{1}\indA{2}}}{(1-\tau)^2}
 +\frac{\dot\alpha_{\indA{1}\indA{2}\indA{3}}
 \dot\alpha_{\indA{1}\indA{2}\indA{3}}}{3(1+\tau)^2}\right)
\nonumber\\
&+\frac{3\hat\mu^{16/3}(1-\tau)^4}
 {2\tau^{8/3}(1+\tau^2)^{1/3}}
 \left[(\tau^2-14\tau+1)
 \,\sigma_{\indI{1}\indI{2}}\sigma_{\indI{1}\indI{2}}
 +(1-\tau)^2\sigma_{\indA{1}\indA{2}}\,\sigma_{\indA{1}\indA{2}}\right]
\nonumber\\
&+\frac{\hat\mu^4(1-\tau)^2}{32\tau^2(1+\tau^2)}
 \left[3(\tau^4-24\tau^3-2\tau^2-24\tau+1)
\, \alpha_{\indI{1}\indA{1}\indA{2}}
 \alpha_{\indI{1}\indA{1}\indA{2}}
 +(1-\tau)^4
 \,\alpha_{\indA{1}\indA{2}\indA{3}}
 \alpha_{\indA{1}\indA{2}\indA{3}}\right],
 \label{eq:so37-decoupled-action}
\end{align}
from which we find their linearized equations of motion. It turns out, that all these equations admit exact analytic solutions, given by
\begin{align}
{\bf (5,1)}\,:&\quad 
\sigma_{\indI{1}\indI{2}}=
\hat\mu^{-2/3}b^{(2)}_{\indI{1}\indI{2}}\,\frac{\tau^{1/3}(1+\tau^2)^{2/3}
(\tau^2+10\tau+1)}{(1-\tau)^4}
+\hat\mu^{-2}b^{(6)}_{\indI{1}\indI{2}}\,\frac{\tau\,(5\tau^2+2\tau+5)}{5(1-\tau)^4}\,,\nonumber\\
{\bf (1,27)}\,:&\quad
\sigma_{\indA{1}\indA{2}}=
\hat\mu^{-2/3}b^{(2)}_{\indA{1}\indA{2}}\,\frac{\tau^{1/3}(1+\tau^2)^{2/3}}{(1-\tau)^2}
+\hat\mu^{-2}b^{(6)}_{\indA{1}\indA{2}}\frac{\tau}{(1-\tau)^2}\,,\nonumber\\
{\bf (3,21)}\,:&\quad
\alpha_{\indI{1}\indA{1}\indA{2}}=
\hat\mu^{-1/3}a^{(1)}_{\indI{1}\indA{1}\indA{2}}\,\frac{\tau^{1/6}(1+\tau+\tau^2)}
{(1-\tau)^2(1+\tau^2)^{1/6}}
+\hat\mu^{-1}a^{(3)}_{\indI{1}\indA{1}\indA{2}}\,\frac{\tau^{1/2}(1+\tau^2)^{1/2}}
{(1-\tau)^2}\,,\nonumber\\
{\bf (1,35)}\,:&\quad
\alpha_{\indA{1}\indA{2}\indA{3}}=
\hat\mu^{-1/3}a^{(1)}_{\indA{1}\indA{2}\indA{3}}\,\frac{\tau^{1/6}(1-\tau+\tau^2)}
{(1-\tau)^2(1+\tau^2)^{1/6}}
+\hat\mu^{-1}a^{(3)}_{\indA{1}\indA{2}\indA{3}}\,\frac{\tau^{1/2}(1+\tau^2)^{1/2}}
{(1-\tau)^2}\,,
 \label{eq:so37-decoupled-solutions}
\end{align}
in terms of eight independent integration constants. As expected, the UV asymptotic powers and the orders at which the independent integration constants enter, agree with those of the fluctuations around the undeformed ${\rm SO}(10)$-invariant background in \eqref{eq:so10series}. To allow a direct comparison, we have introduced in \eqref{eq:so37-decoupled-solutions} the appropriate powers of $\hat\mu$, such that the free coefficients carry the dimensions corresponding to an expansion in the dimensionful coordinate $t$, rather than the dimensionless coordinate $\tau$, c.f.\ \eqref{eq:rhomt}.

Next, we consider the gauge sector carrying the vector field fluctuations. To quadratic order in the fluctuations, the Lagrangian expanded around the polarized background does not carry the modes $\beta_{\indI{1}\indI{2}}$, $\beta_{\indA{1}\indA{2}}$, as expected from the unbroken  ${\rm SO}(3)\times{\rm SO}(7)$ gauge symmetry. The mixed representation $\beta_{\indI{1}\indA{1}}$ in the ${\bf (3,7)}$ couples to the scalar and axion modes in the same representation. The resulting Lagrangian reads
\begin{align}
\mathrm g^{-1}\mathcal L^{(2)}_{(\mathbf3,\mathbf7)}
={}&-16\dot s_{\indI{1}\indA{1}}\dot s_{\indI{1}\indA{1}}
-\frac{4\tau^{2/3}(1+\tau)^2(1+\tau^2)^{1/3}}
 {\hat\mu^{4/3}(1-\tau)^4}
 \dot z_{\indI{1}\indA{1}}\dot z_{\indI{1}\indA{1}}
 +\frac{16\hat\mu^2(1-\tau)(1-\tau+\tau^2)}
 {\tau^{1/2}(1+\tau)(1+\tau^2)^{1/2}}
 s_{\indI{1}\indA{1}}\dot z_{\indI{1}\indA{1}}
 \nonumber\\
&-\frac{64\tau}{1-\tau^2}
 \dot s_{\indI{1}\indA{1}}\beta_{\indI{1}\indA{1}}
+\frac{64\tau^{5/6}(1+\tau^2)^{1/6}
 (1+\tau+\tau^2)}
 {\hat\mu^{2/3}(1-\tau)^4}
 \dot z_{\indI{1}\indA{1}}\beta_{\indI{1}\indA{1}}\
 -\frac{64\tau(4+\tau+4\tau^2)}{(1-\tau)^4}
 \beta_{\indI{1}\indA{1}}\beta_{\indI{1}\indA{1}}\nonumber\\
&-\frac{16\hat\mu^{8/3}(1-\tau)(5+2\tau^2+5\tau^4)}
 {\tau^{1/3}(1+\tau)^3(1+\tau^2)^{2/3}}
 s_{\indI{1}\indA{1}}\beta_{\indI{1}\indA{1}}
-\frac{8\hat\mu^2(1-\tau+\tau^2)}
 {\tau^{1/2}(1+\tau^2)^{1/2}}
 z_{\indI{1}\indA{1}}\beta_{\indI{1}\indA{1}}\nonumber\\
&
-\frac{2\hat\mu^{14/3}(1-\tau)^3(1+\tau)(1+\tau+\tau^2)}
 {\tau^{11/6}(1+\tau^2)^{7/6}}
 s_{\indI{1}\indA{1}}z_{\indI{1}\indA{1}}
 +\frac{\hat\mu^4(1-\tau)^2(1+\tau)^2
 (3-10\tau+3\tau^2)}
 {16\tau^2(1+\tau^2)}
 z_{\indI{1}\indA{1}}z_{\indI{1}\indA{1}}
\nonumber\\
&
+\frac{3\hat\mu^{16/3}(1-\tau)^4
 (1-4\tau-13\tau^2-40\tau^3-13\tau^4-4\tau^5+\tau^6)}
 {\tau^{8/3}(1+\tau)^4(1+\tau^2)^{1/3}}
 s_{\indI{1}\indA{1}}s_{\indI{1}\indA{1}} \,.
 \label{eq:so37-37}
\end{align}
The linearized $\mathrm{SO}(10)$ transformations, which are broken by the background survive as shift transformations with parameter $\Lambda_{\indI{1}\indA{1}}$
\begin{equation}
\delta \beta_{\indI{1}\indA{1}}=\dot\Lambda_{\indI{1}\indA{1}}\,,\quad
\delta s_{\indI{1}\indA{1}}=
 -\frac{2\mathrm g\tau}{1-\tau^2}\,\Lambda_{\indI{1}\indA{1}}\,,\quad
\delta z_{\indI{1}\indA{1}}=
 \frac{8\mathrm g\,\hat\mu^{2/3}\tau^{1/6}(1+\tau+\tau^2)}
 {(1+\tau)^2(1+\tau^2)^{1/6}}\,\Lambda_{\indI{1}\indA{1}}\,,
\label{eq:so37-mixed-gauge-transformations}
\end{equation}
under which the Lagrangian \eqref{eq:so37-37} is invariant. We may thus eliminate the auxiliary field $\beta_{\indI{1}\indA{1}}$ by its algebraic field equation, and fix the gauge symmetry \eqref{eq:so37-mixed-gauge-transformations} by imposing $z_{\indI{1}\indA{1}}=0$\,. 
Consequently, when we study supersymmetry on the gauge-fixed fields, each supersymmetry transformation must be accompanied by a compensating ${\rm SO}(10)$ rotation \eqref{eq:so37-mixed-gauge-transformations} that restores this gauge.
The field equations for the remaining mode $s_{\indI{1}\indA{1}}$ admit the complete analytic solution 
\begin{equation}
{\bf (3,7)}\,:\quad
s_{\indI{1}\indA{1}}=\frac{b^{(2)}_{\indI{1}\indA{1}}}{\hat\mu^{2/3}}\,
\frac{\tau^{1/3}(1+\tau)(\tau+2)(2\tau+1)
(1+\tau^2)^{2/3}}
{2(1-\tau)^3(1+\tau+\tau^2)}
+\frac{b^{(6)}_{\indI{1}\indA{1}}}{\hat\mu^2}\,
\frac{\tau\,(1+\tau)(5\tau^2-\tau+5)}
{5(1-\tau)^3(1+\tau+\tau^2)}
\,,
\label{eq:so37-mixed-solution}
\end{equation}
in terms of two integration constants. We emphasize that compared to the undeformed ${\rm SO}(10)$ case, the gauge symmetry \eqref{eq:so37-mixed-gauge-transformations} halves the number of physical degrees of freedom in this sector. The mode removed by gauge fixing is the one-dimensional analogue of a Goldstone scalar absorbed by a gauge field. On the boundary, the same reduction is expressed by the Ward identity \eqref{eq:Ward}, which relates operators that become dependent after symmetry breaking.

It remains to work out the fluctuation equations in the ${\bf (1,1)}$ singlet sector which mixes the physical modes $u$ and $a$ from the scalars with the dilaton and the einbein fluctuations. We have collected the resulting rather lengthy quadratic Lagrangian in Appendix \ref{subsec:singlet}.
Under diffeomorphisms, the different singlet modes transform as
\begin{align}
 \delta_\xi n&=\dot\xi\,,
\nonumber\\
\delta_\xi v&=-\frac{\hat\mu^{8/3}(1-\tau)^2
 \bigl[5(1+\tau^4)+12\tau(1+\tau^2)+20\tau^2\bigr]}
 {20\tau^{4/3}(1+\tau)^2(1+\tau^2)^{2/3}}\,\xi\,,
 \nonumber\\
\delta_\xi u&=\frac{63\hat\mu^{8/3}(1-\tau)^2(1+\tau^2)^{1/3}}
 {20\tau^{1/3}(1+\tau)^2}\,\xi\,,
\nonumber\\
\delta_\xi a&=\frac{\hat\mu^{10/3}(1-\tau)^6(1-\tau+\tau^2)}
 {\tau^{7/6}(1+\tau)^4(1+\tau^2)^{5/6}}\,\xi\,,
\label{eq:so37-singlet-diffeomorphisms}
\end{align}
under which the Lagrangian \eqref{eq:so37-singlet-component-action} is invariant up to total derivative. We may fix this gauge freedom by setting $v=0$\,. Since the lapse mode $n$ appears as an auxiliary field, we may use the resulting lapse constraint in order to eliminate this mode from the fluctuation equations. We are then left with two second order differential equations for the remaining modes $u$ and $a$. They can be analytically solved as
\begin{align}
u(\tau)={}&
\frac{\tau^{1/3}}
{12\hat\mu^{2/3}(1-\tau)^2(1+\tau^2)^{1/3}\mathcal D(\tau)}
\Bigl[6b^{(2)}\,\bigl(10(1+\tau^6)+108\tau(1+\tau^4)
-13\tau^2(1+\tau^2)+174\tau^3\bigr)\nonumber\\
&\hspace{52mm}{}-\frac{21a^{(3)}}{\hat\mu}\tau
\bigl(-8(1+\tau^4)+3\tau(1+\tau^2)-14\tau^2\bigr)\Bigr]
\nonumber\\
&+\frac{\tau}{5\hat\mu^2(1-\tau)^2\mathcal D(\tau)}
\Bigl[b^{(6)}\,\bigl(25(1+\tau^4)-10\tau(1+\tau^2)+30\tau^2\bigr)
-21\hat\mu a^{(1)}\tau
\bigl(2(1+\tau^2)-\tau\bigr)\Bigr]\,,
\label{eq:so37-singlet-general-u}
\end{align}
and
\begin{align}
a(\tau)={}&
\frac{\tau^{1/6}\,\hat\mu^{-1/3}}
{5(1+\tau)^2(1+\tau^2)^{1/6}\mathcal D(\tau)}
\Bigl[a^{(1)}\,\bigl(25(1+\tau^6)+85\tau(1+\tau^4)
+257\tau^2(1+\tau^2)+184\tau^3\bigr)
\nonumber\\
&\hspace{48mm}{}
-\frac{40}{3\hat\mu}b^{(6)}\,\tau\bigl(5(1+\tau^4)
+3\tau(1+\tau^2)+11\tau^2\bigr)\Bigr]
\nonumber\\
&
{}
\hspace{-2em}
+\frac{\tau^{1/2}\,\hat\mu^{-1}}
{(1+\tau)^2(1+\tau^2)^{3/2}\mathcal D(\tau)}
\Bigl[a^{(3)}\,\bigl(5(1+\tau^8)-17\tau(1+\tau^6)
+27\tau^2(1+\tau^4)
-55\tau^3(1+\tau^2)+44\tau^4\bigr)
\nonumber\\
&\hspace{38mm}{}
-6\hat\mu b^{(2)}\,\tau\bigl(17(1+\tau^6)
+3\tau(1+\tau^4)
+55\tau^2(1+\tau^2)+6\tau^3\bigr)\Bigr]
\,,
\label{eq:so37-singlet-general-a}
\end{align}
in terms of 4 integration constants
$\{b^{(2)}, b^{(6)}, a^{(1)}, a^{(3)} \}$\,. Here, the denominators carry the polynomial
\begin{equation}
 \mathcal D(\tau):= 5\tau^4+12\tau^3+20\tau^2+12\tau+5\,.
 \label{eq:so37-singlet-denominator}
\end{equation}

We have thus worked out the full set of bosonic fluctuations around the polarized background, identified the physical degrees of freedom, and given explicit analytic solutions to all fluctuation equations. Before combining this with supersymmetry, we first need to extend the analysis to the fermionic sector.

\subsubsection*{Fermionic fluctuations}

The polarized background introduces non-trivial Yukawa tensors \eqref{eq:YukawaAll-new}, \eqref{eq:YukA}, triggered by the scalar tensors \eqref{eq:b-objs}
\begin{align}
 \mathring b=\,& 2\,(1-\tau)(5+4\tau+5\tau^2)\,\mathcal{N}(\tau)\,,\quad
 \mathring b^{\indI{1}\indI{2}}=-\frac{14}{5}\,\tau(1-\tau)\,\mathcal N(\tau)\,\delta^{\indI{1}\indI{2}}\,,
 \quad
 \mathring b^{\indA{1}\indA{2}}=\frac{6}{5}\,\tau(1-\tau)\,\mathcal N(\tau)\,\delta^{\indA{1}\indA{2}}
 \,,\nonumber\\
 \mathring b^{\indI{1}\indI{2}\indI{3}}=\,&  8\,\tau^{1/2}(1+\tau+\tau^2)\,\mathcal{N}(\tau)\,\epsilon^{\indI{1}\indI{2}\indI{3}}\,,
 \qquad
 \mathcal{N}(\tau):=
 \frac{\hat\mu^{8/3}\,(1-\tau)^2}{\tau^{4/3}(1+\tau)(1+\tau^2)^{2/3}}\,.
 \label{eq:Yukawa37}
\end{align}
These enter the fermionic fluctuation equations linearly.
Gravitino and dilatino remain in irreducible representations \eqref{eq:1682} under ${\rm SO}(3)\times{\rm SO}(7)$\,. We parametrize the matter fermions as
\begin{equation}
 \chi_{\indI{1}}=\Sigma_{\indI{1}}-\frac13\,\Gamma_{\indI{1}}\Theta\,,\qquad
 \chi_{\indA{1}}=\Sigma_{\indA{1}}+\frac17\,\Gamma_{\indA{1}}\Theta\,,
 \qquad \Gamma^{\indI{1}}\Sigma_{\indI{1}}=\Gamma^{\indA{1}}\Sigma_{\indA{1}}=0\,,
 \label{eq:so37-fermion-decomposition}
\end{equation}
where the $\Sigma_{\indI{1}}$, $\Sigma_{\indA{1}}$ and $\Theta$ carry the irreducible
$(\mathbf4,\mathbf8)$, $(\mathbf2,\mathbf{48})$, and $(\mathbf2,\mathbf8)$, respectively. 
In this normalization,
$\Theta=\Gamma^{\indA{1}}\chi_{\indA{1}}=-\Gamma^{\indI{1}}\chi_{\indI{1}}$.
In particular, $\Gamma^{\indI{1}}\Sigma_{\indI{1}}=\Gamma^{\indA{1}}\Sigma_{\indA{1}}=0$\,.
The trace representation $(\mathbf2,\mathbf8)$ mixes with the dilatino and gravitino fluctuations and transforms under local supersymmetry.

The quadratic Lagrangians for the $\Gamma$-traceless fermions in the $(\mathbf4,\mathbf8)$ and $(\mathbf2,\mathbf{48})$ take the explicit form
\begin{align}
\mathcal L_{(\mathbf4,\mathbf8)}^{(2)}={}&
2\bar\Sigma_{\indI{1}}\dot\Sigma_{\indI{1}}
+\frac{\mathcal{N}(\tau)}
{8\,(1+\tau)}\,
\bar\Sigma_{\indI{1}}\,\Bigl[
(1-\tau^2)(1+12\tau+\tau^2)\Gamma_*
+4\tau^{1/2}(1-\tau)(1-\tau+\tau^2)\Gamma_{\#}
\nonumber\\
&\hspace{40mm}{}
+20\tau^{1/2}(1+\tau)(1+\tau+\tau^2)
\Gamma_{\#}\Gamma_*\Bigr]\,\Sigma_{\indI{1}}\,,
\label{eq:so37-Lag3}\\
\mathcal L_{(\mathbf2,\mathbf{48})}^{(2)}={}&
2\bar\Sigma_{\indA{1}}\dot\Sigma_{\indA{1}}
+\frac{\mathcal{N}(\tau)}
{8\,(1+\tau)}\,
\bar\Sigma_{\indA{1}}\,\Bigl[
(1-\tau^2)(1-4\tau+\tau^2)\Gamma_*
-4\tau^{1/2}(1-\tau)(1-\tau+\tau^2)\Gamma_{\#}
\nonumber\\
&\hspace{40mm}{}
+12\tau^{1/2}(1+\tau)(1+\tau+\tau^2)
\Gamma_{\#}\Gamma_*\Bigr]\,\Sigma_{\indA{1}}\,,
\label{eq:so37-Lag7}
\end{align}
where $\mathcal{N}(\tau)$ is the common factor of the Yukawa tensors \eqref{eq:Yukawa37}.
Since these representations appear just once in the spectrum, the resulting field equations are automatically diagonal. We can give their general solution in analytic form.
Denote by $\sigma_{\pm\,\indI{1}}$ and $\sigma_{\pm\,\indA{1}}$ the constant
gamma-traceless vector-spinors with chirality $\pm$, \textit{i.e.}\
\begin{equation}
\Gamma^{\indI{1}}\sigma_{\pm\,\indI{1}}=\Gamma^{\indA{1}}\sigma_{\pm\,\indA{1}}=0\,,\qquad
\Gamma_* \sigma_{\pm\,\indI{1}} = \pm \sigma_{\pm\,\indI{1}}\,,\quad
\Gamma_* \sigma_{\pm\,\indA{1}} = \pm \sigma_{\pm\,\indA{1}}\,.
\end{equation}
The general solution to the fluctuation equations of \eqref{eq:so37-Lag3} and \eqref{eq:so37-Lag7} is then given by
\begin{align}
\Sigma_{\indI{1}}={}&
\frac{\tau^{-1/12}(1+\tau^2)^{1/3}}
{(1-\tau)^{5/2}(1+\tau)^{1/2}}
\Bigl[(1+2\tau)(1+3\tau)
-\tau^{1/2}(2+\tau)(3+\tau)
\Gamma_{\#}\Bigr]\,\hat\mu^{1/6}\sigma_{+\,\indI{1}}
\nonumber\\
&+
\frac{\tau^{1/12}}
{5(1-\tau)^{5/2}(1+\tau)^{1/2}(1+\tau^2)^{1/3}}
\Bigl[(5+15\tau+6\tau^2+10\tau^3)
\nonumber\\
&\hspace{62mm}{}+\tau^{1/2}(10+6\tau+15\tau^2+5\tau^3)
\Gamma_{\#}\Bigr]\,\hat\mu^{-1/6}\sigma_{-\,\indI{1}}\,,
\nonumber\\[1ex]
\Sigma_{\indA{1}}={}&
\frac{\tau^{-1/12}(1+\tau^2)^{1/3}}
{(1-\tau)^{3/2}(1+\tau)^{1/2}}
\Bigl[(1+2\tau)
-\tau^{1/2}(2+\tau)\Gamma_{\#}\Bigr]\,\hat\mu^{1/6}\sigma_{+\,\indA{1}}
\nonumber\\
&+
\frac{\tau^{1/12}}
{(1-\tau)^{3/2}(1+\tau)^{1/2}(1+\tau^2)^{1/3}}
\Bigl[(1+2\tau^2)
+\tau^{1/2}(2+\tau^2)\Gamma_{\#}\Bigr]\,\hat\mu^{-1/6}\sigma_{-\,\indA{1}} \,,
\label{eq:fermions-sol}
\end{align}
in terms of the 4 constant spinors.
Just as for the scalar fluctuations, the UV asymptotic powers and the orders at which the independent integration constants enter agree with those of the fermionic fluctuations around the undeformed ${\rm SO}(10)$-invariant background in \eqref{eq:so10chiseries}.
The spinors \eqref{eq:fermions-sol} can be obtained by applying the background supersymmetry to the explicit bosonic solutions \eqref{eq:so37-decoupled-solutions}, as we illustrate in the next section. Substitution into the first-order fluctuation equations from \eqref{eq:so37-Lag3} and \eqref{eq:so37-Lag7} then provides an independent check.

We are left with the fermionic fluctuations in the fundamental ${\bf (2,8)}$ representation. The modes come from the gravitino, the dilatino, and the trace part $\Theta$ of the matter fermions \eqref{eq:so37-fermion-decomposition}. Their quadratic Lagrangian can be worked out explicitly and is collected in Appendix~\ref{subsec:spinor}. The further reduction of this sector follows the same interplay of supersymmetry, gauge fixing, and field equations as in the ${\rm SO}(10)$ case above, though the calculations are considerably more involved. We first note that the gravitino field equations again imply an algebraic constraint which can be solved to express the positive chirality part of the dilatino $\lambda_+$ in terms of the other fields. Contrary to the ${\rm SO}(10)$ case both gravitino chiralities $\psi_+$ and $\psi_-$ enter the Lagrangian, however in the precise combination
\begin{equation}
\psi_-+\tau^{1/2}\,\Gpolar\,\psi_+\,,
\label{eq:gravitino-comb}
\end{equation}
such that the associated constraint only allows to eliminate one chirality of the dilatino.

Next, we spell out the linearized supersymmetry transformations in this sector, which are given by
\begin{align}
\delta_\eta\lambda={}&
\frac{\mathrm g \hat\mu^{8/3}(1-\tau)^2}
 {5\tau^{4/3}(1+\tau)^2(1+\tau^2)^{2/3}}
\Bigl[(1-\tau^2)(5+4\tau+5\tau^2)
-(5\tau^4+12\tau^3+20\tau^2+12\tau+5)\,\Gamma_*\nonumber\\
&\hspace{25mm}
+9\tau^{1/2}(1+\tau)(1+\tau+\tau^2)\Gamma_{\#}
+\tau^{1/2}(1-\tau)(1-\tau+\tau^2)
 \Gamma_{\#}\Gamma_*\Bigr]\eta\,,
\nonumber\\
\delta_\eta\Theta={}&
-\frac{21\mathrm g \hat\mu^{8/3}(1-\tau)^2}
 {5\tau^{4/3}(1+\tau)^2(1+\tau^2)^{2/3}}
\Bigl[-3\tau(1+\tau^2)+\tau(1-\tau^2)\Gamma_*\nonumber\\
&\hspace{25mm}
+\tau^{1/2}(1-\tau)(1-\tau+\tau^2)\Gamma_{\#}
-\tau^{1/2}(1+\tau)(1+\tau+\tau^2)
 \Gamma_{\#}\Gamma_*\Bigr]\eta\,,
\nonumber\\
\delta_\eta\psi={}&\dot\eta+
\frac{\hat\mu^{8/3}(1-\tau)^2}
 {16\tau^{4/3}(1+\tau)^2(1+\tau^2)^{2/3}}
\Bigl[
4\tau^{1/2}(1-\tau)(1-\tau+\tau^2)\Gamma_{\#}
\nonumber\\
&\hspace{25mm}
-(1-\tau^2)(5+4\tau+5\tau^2)\Gamma_*
+12\tau^{1/2}(1+\tau)(1+\tau+\tau^2)
 \Gamma_{\#}\Gamma_*
\Bigr]\eta \,,
\label{eq:so37-28-local-susy}
\end{align}
for an arbitrary local-supersymmetry parameter $\eta(t)$\,. As in the ${\rm SO}(10)$ case, we use local $\eta_-$ supersymmetry and the resulting
\begin{equation}
\delta_{\eta_-}\lambda_-=
\frac{2\mathrm g \hat\mu^{8/3}(1-\tau)^2
 (4\tau^3+10\tau^2+8\tau+5)}
 {5\tau^{4/3}(1+\tau)^2(1+\tau^2)^{2/3}}\,
\eta_- \,,
\label{eq:so37-28-gauge-shift}
\end{equation}
in order to impose the gauge $\lambda_-=0$.  A subsequent global supersymmetry transformation
must be accompanied by the compensating $\eta_-$ that restores this gauge.  The gravitino constraint then fixes the remaining dilatino algebraically in terms of $\Theta$, while the dilatino equation determines the gravitino \eqref{eq:gravitino-comb}. Put together, this reduces the ${\bf (2,8)}$ sector to a first order differential equation for the trace mode $\Theta$, which can be explicitly solved as
\begin{align}
\Theta(\tau)={}&-\frac{\tau^{1/12}(2\tau^2+1)
 (8\tau^3-50\tau^2+16\tau-25)}
 {5Q(\tau)(1+\tau^2)^{1/3}}\,\hat\mu^{-1/6}\theta_+
\nonumber\\
&
+\frac{2\tau^{5/12}(1+\tau^2)^{1/3}
 (2\tau^4+2\tau^3+28\tau^2+7\tau+12)}
 {Q(\tau)}\,
 \Gamma_{\#}\hat\mu^{1/6}\theta_-
 \nonumber\\
&
+\frac{5(1+\tau^2)^{1/3}
 (3\tau^4+2\tau^3+8\tau^2-2\tau+1)}
 {\tau^{1/12}\,Q(\tau)}\,\hat\mu^{1/6}\theta_-\nonumber\\
&-\frac{2\tau^{7/12}(\tau^2+2)
 (2\tau^3-2\tau^2+4\tau-1)}
 {Q(\tau)(1+\tau^2)^{1/3}}\,
 \Gamma_{\#}\hat\mu^{-1/6}\theta_+.
\label{eq:so37-28-solution-minus}
\end{align}
in terms of constant spinors $\theta_\pm$ of positive and negative chirality, respectively, and with the overall denominator
\begin{equation}
Q(\tau):=(1-\tau)^{3/2}(1+\tau)^{1/2}
       (4\tau^3+10\tau^2+8\tau+5)\,.
\label{eq:so37-28-solution-denominator}
\end{equation}
Thus $\Theta_-\sim\tau^{-1/12}\theta_-$ and $\Theta_+\sim\tau^{1/12}\theta_+$ at the ultraviolet boundary, just as for the other fermions \eqref{eq:fermions-sol} and the ${\rm SO}(10)$ case.
This completes our discussion of the fluctuation equations.


\section{Supermultiplets of sources}

\label{sec:supermultiplets}


Having spelled out the asymptotic expansions of the solutions to the fluctuation equations for all the modes, we can now study the induced action of a supersymmetry transformation on the asymptotic data parametrizing the solutions. To this end, we plug the explicit solutions into the linearized supersymmetry transformation rules and evaluate order by order in the asymptotic expansion. Again, we first treat the simpler case of the undeformed ${\rm SO}(10)$ invariant background before discussing the polarized ${\rm SO}(3)\times{\rm SO}(7)$ invariant domain wall. We determine the resulting supermultiplet structure of the sources and find a precise match with the supermultiplets of BPS operators in the undeformed and the polarized IKKT model identified in sections~\ref{subsec:IKKT} and \ref{subsec:F4} above.

\subsection{${\rm SO}(10)$ BPS domain wall}

Insertion of the explicit solutions 
\eqref{eq:so10series}, \eqref{eq:so10chiseries} together with the Killing spinor \eqref{eq:solSO10} 
into the linearized supersymmetry rules allows to translate the supersymmetry action into an action on the sources parametrizing the solutions.
In the normalization inherited directly from the bulk fields, we find that under supersymmetry the sources transform within two closed subsets
\begin{align}
\delta b^{(2)}_{ab}
 &=\bar\epsilon_0\Gamma^{\vphantom{+}}_{(a}\kappa^+_{b)}\,,&
\delta\kappa^+_a
 &=\frac{\mathrm g}{4}\,\mathbb{P}_a{}^d a^{(3)}_{dbc}\,\Gamma^{bc}\epsilon_0\,,&
\delta a^{(3)}_{abc}&=0\,,
\nonumber\\
\delta a^{(1)}_{abc}
 &=-3\bar\epsilon_0\Gamma^{\vphantom{-}}_{[ab}\kappa^-_{c]}\,,&
\delta\kappa^-_a
 &=2\mathrm g\, b^{(6)}_{ab}\,\Gamma^b\epsilon_0\,,&
\delta b^{(6)}_{ab}&=0\,,
\label{eq:so10raw2}
\end{align}
where we have defined the gamma-traceless projector
\begin{equation}
 \mathbb{P}_a{}^b=\delta_a{}^b-\frac1{10}\Gamma_a\Gamma^b\,.
\end{equation}
The $b^{(2)}$ and $a^{(1)}$ terms drop out separately in $\delta\chi_a$. These
cancellations correspond to the linearized BPS equations $\delta_\eps \chi_a=0$\,.

Setting
\begin{equation}
 B_{ab}=b^{(2)}_{ab}\,,\qquad
 \Psi_a=\frac59\,\kappa^+_a\,,\qquad
 A_{abc}=\frac{\mathrm g}{4}\,a^{(3)}_{abc}\,,
 \label{eq:rescale-M1}
\end{equation}
the first line of \eqref{eq:so10raw2} reads
\begin{equation}
\delta B_{ab}
 =\frac95\,\bar\epsilon_0\Gamma_{(a}\Psi_{b)}\,,\qquad
 \delta\Psi^a
 =\frac1{18}\left(
 7A^{abc}\Gamma_{bc}-A_{bcd}\Gamma^{abcd}\right)\epsilon_0\,,\qquad
\delta A_{abc}=0\,,
\label{eq:so10canonical1}
\end{equation}
and precisely reproduces the transformations of the IKKT BPS multiplet \eqref{eq:OOOsusy}
\begin{equation}
\overset{\raisebox{1ex}{$\scriptstyle B_{ab}$}}{\mathbf{54}}
\longrightarrow\overset{\raisebox{1ex}{$\scriptstyle \Psi_a$}}{\mathbf{144}}
\longrightarrow\overset{\raisebox{1ex}{$\scriptstyle A_{abc}$}}{\mathbf{120}}
\,.
 \label{eq:54-120}
\end{equation}
Similarly, after rescaling the sources in the second line of \eqref{eq:so10raw2}
\begin{equation}
 \widetilde A_{abc}=a^{(1)}_{abc}\,,\qquad
 \widetilde\Psi_a=-\frac{27}{5}\kappa^-_a\,,\qquad
 \widetilde B_{ab}=-6\mathrm g b^{(6)}_{ab}\,,
 \label{eq:rescale-M2}
\end{equation}
their transformations take the form
\begin{equation}
\delta\widetilde A_{abc}
 =\frac59\bar\epsilon_0\Gamma_{[ab}\widetilde\Psi_{c]}\,,\qquad
\delta\widetilde\Psi_a
 =\frac95\widetilde B_{ab}\Gamma^b\epsilon_0\,,\qquad
\delta\widetilde B_{ab}=0\,.
\label{eq:so10canonical2}
\end{equation}
This describes a supermultiplet dual to \eqref{eq:54-120}
\begin{equation}
\overset{\raisebox{1ex}{$\scriptstyle \tilde{A}_{abc}$}}{\mathbf{120}}
\longrightarrow\overset{\raisebox{1ex}{$\scriptstyle \tilde{\Psi}_a$}}{\mathbf{144}}
\longrightarrow
\overset{\raisebox{1ex}{$\scriptstyle \tilde{B}_{ab}$}}{\mathbf{54}}
\,,
 \label{eq:120-54}
\end{equation}
such that the bilinear pairing
\begin{equation}
 \mathcal I
 =A_{abc}\widetilde A^{abc}
  +\bar\Psi^{a}\widetilde\Psi_a
  +B_{ab}\widetilde B^{ab}\,,
  \label{eq:so10-dual-pairing}
\end{equation}
is invariant under supersymmetry.
The two branches therefore contain the same three ${\rm SO}(10)$ representations, but are
oppositely oriented under the nilpotent supersymmetry. They form the
short multiplet ${\cal B}_2$ and its dual.
It is straightforward to verify that the supersymmetry transformations \eqref{eq:so10canonical1} and \eqref{eq:so10canonical2} close into an abelian algebra.

In view of a holographic description of the lowest IKKT multiplet \eqref{eq:OOOsusy}, we conclude that the relevant sources should transform in the dual multiplet \eqref{eq:120-54}, \textit{i.e.}\ the sources are given by the leading term of the axion fields $a_{abc}$, the negative chirality part of the fermions $\chi^a$, together with the subleading term of the scalars ${\cal T}_{ab}$.
In the spirit of \cite{Klebanov:1999tb}, this would likely require a Legendre transform of the holographically renormalized on-shell action in the ${\bf 54}$ sector. We leave a more detailed analysis to future work.


\subsection{${\rm SO}(3)\times {\rm SO}(7)$ BPS domain wall}

We now extend this computation to the polarized ${\rm SO}(3)\times {\rm SO}(7)$ background. Compared to the ${\rm SO}(10)$ case, the computation is more involved due to the larger number of irreducible components and the more complicated coordinate dependence of the on-shell fluctuations. The method however remains exactly the same.
We insert the on-shell fluctuations and the Killing spinor \eqref{eq:BPSpolKS} into the supersymmetry rules \eqref{eq:SUSYbos-new}, \eqref{eq:SUSYfer-new}, of the bulk theory, including compensating transformations where required by our gauge choices. Equating the two sides of the variation determines how each bosonic integration constant varies into fermionic constants, and conversely.
This establishes the supermultiplet structure of the asymptotic sources. At the same time, the computation tests more than just the leading asymptotic powers: after all terms are combined, each variation must reproduce the entire $\tau$-dependent profile of an on-shell mode with a constant coefficient. In practice, we have even used this computation in order to find the explicit form of the fermionic fluctuations \eqref{eq:fermions-sol} and \eqref{eq:so37-28-solution-minus} from the known bosonic solutions, rather than trying to integrate their differential equations. We then verified directly that the solutions obtained in this way satisfy the fermionic fluctuation equations.

Let us present this in detail for the supersymmetry variation of the $({\bf2},{\bf48})$
spinor $\Sigma_{\indA{1}}$, describing an irreducible block \eqref{eq:so37-fermion-decomposition} of the matter fermions $\chi^a$. We parametrize this spinor as 
\begin{equation}
 \Sigma_{\indA{1}}(\tau)=
 \zeta_+(\tau)\sigma_{+\,\indA{1}}
 +\zeta_-(\tau)\sigma_{-\,\indA{1}}
 \quad
 \Longrightarrow\quad
\delta  \Sigma_{\indA{1}}(\tau)=
 \zeta_+(\tau)\,\delta\sigma_{+\,\indA{1}}
 +\zeta_-(\tau)\,\delta\sigma_{-\,\indA{1}}
 \,,
 \label{eq:so37-example-unknown-profiles}
\end{equation}
with a priori unknown (Clifford-valued) functions $\zeta_\pm(\tau)$, and $\Gamma$-traceless constant spinors $\sigma_{\pm\,\indA{1}}$ of positive and negative chirality, respectively.
For the right-hand side of the variation, we project the bulk variation \eqref{eq:SUSYfer-new} onto the
seven-dimensional gamma-traceless vector spinor. We then expand the r.h.s.\ to linearized order in the fluctuations upon using the explicit form of the bosonic on-shell fluctuations \eqref{eq:so37-decoupled-solutions}, \eqref{eq:so37-mixed-solution}. Representation-wise, the singlet fluctuations cannot contribute to the variation of the $({\bf2},{\bf48})$, such that we can replace in this expression dilaton and einbein by their background values. This yields
\begin{equation}
\delta_\epsilon\Sigma_{\indA{1}}
= \mathbb{P}_{\indA{1}}{}^{\indA{2}}\,\delta \chi_{\indA{2}}
= {\rm g}\,\mathbb{P}_{\indA{1}\indA{2}}\,
\Big(
4\,P^{\indA{2}b}\,\Gamma_b
 +\frac{7e^{-\mathring\phi}}{10}\, p_{bcd}\,\widetilde\Gamma^{\indA{2}b,cd}
 + C^{\indA{2}}\Big)\, \mathring{\epsilon}
\,,
\label{eq:delSigmaA}
\end{equation}
with the projector from \eqref{eq:proj37} and the explicit Killing spinor from \eqref{eq:BPSpolKS}.
Let us focus on the contribution of the bosonic fluctuation $\sigma_{\indA{1}\indA{2}}$ which enters the r.h.s.\ through
\begin{equation}
P^{\indA{1}\indA{2}}\rightarrow\dot\sigma^{\indA{1}\indA{2}}\,,\qquad
 C^{\indA{1}} \rightarrow -e^{4\mathring\phi}\mathring{X}^{-3/10} \sigma^{\indA{1}\indA{2}}
 \Gamma_{\indA{2}}\Gamma_*\,,
 \label{eq:so37-example-scalar-currents}
\end{equation}
such that we are left with
\begin{equation}
 \left.\delta_\epsilon\Sigma_{\indA{1}}\right|_{\sigma^{\indA{1}\indA{2}}}
 =4\mathrm g\dot\sigma_{\indA{1}\indA{2}}
 \,\Gamma^{\indA{2}}\,\mathring{\epsilon}
 -\mathrm g e^{4\mathring\phi}\mathring{X}^{-3/10}
 \sigma_{\indA{1}\indA{2}}
 \Gamma^{\indA{2}}\Gamma_*\,\mathring{\epsilon}\,.
 \label{eq:so37-example-two-terms}
\end{equation}
The tracelessness and symmetry of $\sigma_{\indA{1}\indA{2}}$ make
$\Gamma^{\indA{2}}\sigma_{\indA{1}\indA{2}}\epsilon$
gamma-traceless, such that the projector acts as an identity here. 
Using the explicit form of $\sigma_{\indA{1}\indA{2}}$, 
given by the second line of \eqref{eq:so37-decoupled-solutions},
we find that for its $b^{(2)}$ branch, the two terms in \eqref{eq:so37-example-two-terms} yield the contributions
\begin{align}
 4\mathrm g\dot\tau\frac{\mathrm d}{\mathrm d\tau}
 \left[\frac{\tau^{1/3}(1+\tau^2)^{2/3}}
 {(1-\tau)^2}\right]
 &=\mathrm g\,\hat\mu^{8/3}\,\frac{1+4\tau+\tau^2}{\tau},\nonumber\\
 \mathrm g e^{4\mathring\phi}\mathring{X}^{-3/10}
 \frac{\tau^{1/3}(1+\tau^2)^{2/3}}
 {(1-\tau)^2}
 &=\mathrm g\,\hat\mu^{8/3}\frac{1-\tau^2}{\tau}.
 \label{eq:so37-example-first-reduction}
\end{align}
Putting everything together with the explicit form \eqref{eq:BPSpolKS} of the Killing spinor, we thus obtain
\begin{align}
 \left.\delta_\epsilon\Sigma_{\indA{1}}\right|_{b^{(2)}}
 &=\frac{\mathrm g\,\hat\mu^{7/6}\,\tau^{-7/12}(1+\tau^2)^{1/3}}
 {(1-\tau)^{3/2}(1+\tau)^{1/2}}
 b^{(2)}_{\indA{1}\indA{2}}\Gamma^{\indA{2}}
 \left[
 (1+4\tau+\tau^2)
 -(1-\tau^2)\Gamma_*\right](1-\tau^{1/2}\Gpolar)\,\epsilon_0
 \,,
 \nonumber\\
 &=\hat\mu^{1/6}\frac{\tau^{-1/12}(1+\tau^2)^{1/3}}
 {(1-\tau)^{3/2}(1+\tau)^{1/2}}
 \left[(1+2\tau)-\tau^{1/2}(2+\tau)\Gpolar\right]
 \left[-2\mathrm g\,\hat\mu b^{(2)}_{\indA{1}\indA{2}}
 \Gamma^{\indA{2}}\Gpolar\epsilon_0\right].
 \label{eq:so37-example-first-factorization}
\end{align}
Similarly, we find for the $b^{(6)}$ branch
\begin{equation}
 \left.\delta_\epsilon\Sigma_{\indA{1}}\right|_{b^{(6)}}
 =\hat\mu^{-1/6}\frac{\tau^{1/12}}
 {(1-\tau)^{3/2}(1+\tau)^{1/2}(1+\tau^2)^{1/3}}
 \left[(1+2\tau^2)+\tau^{1/2}(2+\tau^2)\Gpolar\right]
 \left[2\mathrm g b^{(6)}_{\indA{1}\indA{2}}
 \Gamma^{\indA{2}}\epsilon_0\right].
 \label{eq:so37-example-second-factorization}
\end{equation}
Comparing these expressions to the l.h.s.\ of the variation \eqref{eq:so37-example-unknown-profiles}, we can read off the form of the functions $\zeta_\pm$
\begin{align}
 \zeta_+(\tau)=\,&\hat\mu^{1/6}\,
 \frac{\tau^{-1/12}(1+\tau^2)^{1/3}}
 {(1-\tau)^{3/2}(1+\tau)^{1/2}}
 \,\bigl[(1+2\tau)-\tau^{1/2}(2+\tau)\Gpolar\bigr]\,,
 \nonumber\\
 \zeta_-(\tau)=\,&\hat\mu^{-1/6}\,
 \frac{\tau^{1/12}}
 {(1-\tau)^{3/2}(1+\tau)^{1/2}(1+\tau^2)^{1/3}}
 \,\bigl[(1+2\tau^2)+\tau^{1/2}(2+\tau^2)\Gpolar\bigr]\,,
 \label{eq:so37-example-determined-profiles}
\end{align}
up to some global normalization factors.
We have thus determined the form of the fermionic on-shell fluctuations directly by supersymmetry. As a consistency check, one may verify that the resulting expressions for $\Sigma_{\indA{1}}$ indeed satisfy the linear fluctuation equations obtained from variation of \eqref{eq:so37-Lag7}.
Moreover, comparison of \eqref{eq:so37-example-first-factorization}, \eqref{eq:so37-example-second-factorization} with \eqref{eq:so37-example-unknown-profiles} yields the supersymmetry variation of the sources
\begin{equation}
\delta_\epsilon\sigma_{+\,\indA{1}}\big|_{b^{(2)}}
=-2\mathrm g\,\hat\mu b^{(2)}_{\indA{1}\indA{2}}
\Gamma^{\indA{2}}\Gpolar\epsilon_0\,,\qquad
\delta_\epsilon\sigma_{-\,\indA{1}}\big|_{b^{(6)}}
=2\mathrm g b^{(6)}_{\indA{1}\indA{2}}
\Gamma^{\indA{2}}\epsilon_0\,.
\end{equation}
To complete the result, we must include the other scalar modes on the r.h.s.\ of \eqref{eq:delSigmaA}. Without going through the details of the computation, which closely parallels the preceding one, we simply state the final result for the full variation
\begin{align}
 \delta_\epsilon\sigma_{+\,\indA{1}}
 =\,& \mathbb{P}_{\indA{1}}{}^{\indA{2}}\,
 \Bigg[-\frac{\mathrm g}{2}
 a^{(3)}_{\indI{1}\indA{2}\indA{3}}
 \Gamma^{\indI{1}\indA{3}}
 +\frac{\mathrm g}{2}
 a^{(3)}_{\indA{2}\indA{3}\indA{4}}
 \Gamma^{\indA{3}\indA{4}}
 +\mathrm g \hat\mu
 \left(5b^{(2)}_{\indI{1}\indA{2}}
 \Gamma^{\indI{1}}
 -2b^{(2)}_{\indA{2}\indA{3}}
 \Gamma^{\indA{3}}\right)\Gpolar\Bigg]\epsilon_0\,,
 \nonumber\\[1ex]
 \delta_\epsilon\sigma_{-\,\indA{1}}
 =\,& \mathbb{P}_{\indA{1}}{}^{\indA{2}}\,
 \Bigg[2\mathrm g
 \left(b^{(6)}_{\indI{1}\indA{2}}\Gamma^{\indI{1}}
 +b^{(6)}_{\indA{2}\indA{3}}\Gamma^{\indA{3}}\right)
-\frac{\mathrm g \hat\mu}{2}
 \left(3a^{(1)}_{\indI{1}\indA{2}\indA{3}}
 \Gamma^{\indI{1}\indA{3}}
 +a^{(1)}_{\indA{2}\indA{3}\indA{4}}
 \Gamma^{\indA{3}\indA{4}}\right)
 \Gpolar\Bigg]\,\epsilon_0\,.
 \label{eq:so37-var-fermi1}
\end{align}
Each scalar mode provides another consistency test for the $\tau$-profile \eqref{eq:so37-example-determined-profiles} of the fermionic fluctuations.

A very similar computation yields the supersymmetry transformation of the other fermionic sources as
\begin{align}
\delta_\epsilon\sigma_{+\,\indI{1}}={}&
 \mathbb P_{\indI{1}}{}^{\indI{2}}
 \Bigl[\mathrm g \hat\mu
 \bigl(6b^{(2)}_{\indI{2}\indI{3}}\Gamma^{\indI{3}}
 -3b^{(2)}_{\indI{2}\indA{1}}\Gamma^{\indA{1}}\bigr)
 \Gpolar
 +\frac{\mathrm g}{2}
 a^{(3)}_{\indI{2}\indA{1}\indA{2}}
 \Gamma^{\indA{1}\indA{2}}\Bigr]\,\epsilon_0\,,
 \nonumber\\
\delta_\epsilon\sigma_{-\,\indI{1}}={}&
 \mathbb P_{\indI{1}}{}^{\indI{2}}
 \Bigl[2\mathrm g
 \bigl(b^{(6)}_{\indI{2}\indI{3}}\Gamma^{\indI{3}}
 +b^{(6)}_{\indI{2}\indA{1}}\Gamma^{\indA{1}}\bigr)
 -\frac{5\mathrm g \hat\mu}{2}
 a^{(1)}_{\indI{2}\indA{1}\indA{2}}
 \Gamma^{\indA{1}\indA{2}}\Gpolar\Bigr]\,\epsilon_0\,,
\nonumber\\[2ex]
 \delta_\epsilon\theta_-={}&
 -3\mathrm g \hat\mu
 \left(\frac85b^{(2)}
 +\frac12b^{(2)}_{\indI{1}\indA{1}}
 \Gamma^{\indI{1}\indA{1}}\right)
 \Gpolar\epsilon_0
 -\frac{3\mathrm g}{2}
 \left[\frac1{30}a^{(3)}_{\indI{1}\indA{1}\indA{2}}
 \Gamma^{\indI{1}\indA{1}\indA{2}}
 -\frac3{10}a^{(3)}_{\indA{1}\indA{2}\indA{3}}
 \Gamma^{\indA{1}\indA{2}\indA{3}}
 +\frac7{10}a^{(3)}\Gpolar\right]\epsilon_0\,,
\nonumber\\
 \delta_\epsilon\theta_+={}&
 -3\mathrm g
 \left(-\frac23b^{(6)}+\frac23b^{(6)}_{\indI{1}\indA{1}}
 \Gamma^{\indI{1}\indA{1}}\right)\epsilon_0
-3\mathrm g \hat\mu
 \left[\left(\frac16a^{(1)}_{\indI{1}\indA{1}\indA{2}}
 \Gamma^{\indI{1}\indA{1}\indA{2}}
 +\frac12a^{(1)}_{\indA{1}\indA{2}\indA{3}}
 \Gamma^{\indA{1}\indA{2}\indA{3}}\right)\Gpolar
 +\frac7{10}a^{(1)}\right]\epsilon_0\,.
 \label{eq:var-fermi2}
\end{align}
In this computation, a compensating local supersymmetry transformation has to be taken into account in order to restore the gauge $\lambda_-=0$\,.

Finally, computing the supersymmetry variation of the bosonic sources needs to take into account compensating diffeomorphisms and ${\rm SO}(10)$ rotations. After some straightforward computation, we find for the variation of the constants in the first branch 
\begin{align}
\delta_\epsilon b^{(2)}_{\indI{1}\indI{2}}
 & =\bar\epsilon_0\Gamma_{(\indI{1}}
 \sigma_{+\,\indI{2})}
 -\frac13\delta_{\indI{1}\indI{2}}
 \bar\epsilon_0\Gamma^{\indI{3}}\sigma_{+\,\indI{3}}\,,
 \qquad
\delta_\epsilon b^{(2)}_{\indA{1}\indA{2}}
  =\bar\epsilon_0\Gamma_{(\indA{1}}
 \sigma_{+\,\indA{2})}
 -\frac17\delta_{\indA{1}\indA{2}}
 \bar\epsilon_0\Gamma^{\indA{3}}\sigma_{+\,\indA{3}}\,,
 \nonumber\\
\delta_\epsilon b^{(2)}_{\indI{1}\indA{1}}
 & =\frac12\bar\epsilon_0\Gamma_{\indA{1}}
 \sigma_{+\,\indI{1}}
 +\frac12\bar\epsilon_0\Gamma_{\indI{1}}
 \sigma_{+\,\indA{1}}
 +\frac5{21}\bar\epsilon_0\Gamma_{\indI{1}\indA{1}}\theta_-\,,
 \qquad
\delta_\epsilon b^{(2)}
  =\bar\epsilon_0\theta_-\,,
 \qquad
 \delta_\epsilon a^{(3)}
  =-6\hat\mu\bar\epsilon_0\theta_-\,,
\nonumber\\
\delta_\epsilon a^{(3)}_{\indI{1}\indA{1}\indA{2}}
 & =\hat\mu\Bigl[
 5\bar\epsilon_0\Gamma_{\indA{1}\indA{2}}
 \Gpolar\sigma_{+\,\indI{1}}
 +6\bar\epsilon_0\Gamma_{\indI{1}[\indA{1}}
 \Gpolar\sigma_{+\,\indA{2}]}
 +\frac{10}{21}\bar\epsilon_0
 \Gamma_{\indI{1}\indA{1}\indA{2}}
 \Gpolar\theta_-\Bigr]\,,
 \nonumber\\
\delta_\epsilon a^{(3)}_{\indA{1}\indA{2}\indA{3}}
 & =\hat\mu\Bigl[
 3\bar\epsilon_0\Gamma_{[\indA{1}\indA{2}}
 \Gpolar\sigma_{+\,\indA{3}]}
 -\frac{10}{7}\bar\epsilon_0
 \Gamma_{\indA{1}\indA{2}\indA{3}}
 \Gpolar\theta_-\Bigr]\,.
 \label{eq:so37-asymptotic-bosonic-first}
\end{align}
For the constants on the second branch, we find
\begin{align}
\delta_\epsilon b^{(6)}_{\indI{1}\indI{2}}
 & =\hat\mu\Bigl[3\bar\epsilon_0\Gamma_{(\indI{1}}
 \Gpolar\sigma_{-\,\indI{2})}
 -\delta_{\indI{1}\indI{2}}
 \bar\epsilon_0\Gamma^{\indI{3}}\Gpolar\sigma_{-\,\indI{3}}\Bigr]\,,\qquad
\delta_\epsilon b^{(6)}_{\indA{1}\indA{2}}
  =\hat\mu\Bigl[\bar\epsilon_0\Gamma_{(\indA{1}}
 \Gpolar\sigma_{-\,\indA{2})}
 -\frac17\delta_{\indA{1}\indA{2}}
 \bar\epsilon_0\Gamma^{\indA{3}}\Gpolar\sigma_{-\,\indA{3}}\Bigr]\,,
 \nonumber\\
\delta_\epsilon b^{(6)}_{\indI{1}\indA{1}}
 & =\hat\mu\Bigl[\frac34\bar\epsilon_0\Gamma_{\indA{1}}
 \Gpolar\sigma_{-\,\indI{1}}
 +\frac54\bar\epsilon_0\Gamma_{\indI{1}}
 \Gpolar\sigma_{-\,\indA{1}}
 -\frac5{28}\bar\epsilon_0
 \Gamma_{\indI{1}\indA{1}}\Gpolar\theta_+\Bigr]\,,\qquad
\delta_\epsilon b^{(6)}
  =\frac{9\hat\mu}{5}\bar\epsilon_0\Gpolar\theta_+\,,
 \nonumber\\
\delta_\epsilon a^{(1)}
  & =\bar\epsilon_0\Gpolar\theta_+\,, 
 \qquad
\delta_\epsilon a^{(1)}_{\indI{1}\indA{1}\indA{2}}
  =\Bigl[
 -\bar\epsilon_0\Gamma_{\indA{1}\indA{2}}
 \sigma_{-\,\indI{1}}
 -2\bar\epsilon_0\Gamma_{\indI{1}[\indA{1}}
 \sigma_{-\,\indA{2}]}
 +\frac1{21}\bar\epsilon_0
 \Gamma_{\indI{1}\indA{1}\indA{2}}\theta_+\Bigr]\,,
 \nonumber\\
\delta_\epsilon a^{(1)}_{\indA{1}\indA{2}\indA{3}}
 & =\Bigl[
 -3\bar\epsilon_0\Gamma_{[\indA{1}\indA{2}}
 \sigma_{-\,\indA{3}]}
 -\frac37\bar\epsilon_0
 \Gamma_{\indA{1}\indA{2}\indA{3}}\theta_+\Bigr]
  \,.
 \label{eq:so37-asymptotic-bosonic-second}
\end{align}

\begin{figure}[tb]
\centering
\begin{subfigure}[t]{0.49\textwidth}
\centering
\resizebox{0.91\linewidth}{!}{%
\begin{tikzpicture}[x=1cm,y=1cm,
 every node/.style={draw,rounded corners=2pt,fill=white,inner xsep=5pt,inner ysep=3pt,minimum height=0.65cm},
 every path/.style={thick}]
\node (B) at (-4.5,1.4) {$b^{(2)}_{\indI{1}\indI{2}}\;{\scriptstyle (\mathbf5,\mathbf1)}$};
\node (C) at (-1.5,1.4) {$b^{(2)}_{\indA{1}\indA{2}}\;{\scriptstyle (\mathbf1,\mathbf{27})}$};
\node (R) at (1.5,1.4) {$b^{(2)}_{\indI{1}\indA{1}}\;{\scriptstyle (\mathbf3,\mathbf7)}$};
\node (H) at (4.5,1.4) {$b^{(2)}\;{\scriptstyle (\mathbf1,\mathbf1)}$};
\node (S) at (-3.0,0) {$\sigma_{+\,\indI{1}}\;{\scriptstyle (\mathbf4,\mathbf8)}$};
\node (P) at (0,0) {$\sigma_{+\,\indA{1}}\;{\scriptstyle (\mathbf2,\mathbf{48})}$};
\node (T) at (3.0,0) {$\theta_-\;{\scriptstyle (\mathbf2,\mathbf8)}$};
\node (A) at (1.5,-1.4) {$a^{(3)}_{\indI{1}\indA{1}\indA{2}}\;{\scriptstyle (\mathbf3,\mathbf{21})}$};
\node (D) at (-1.5,-1.4) {$a^{(3)}_{\indA{1}\indA{2}\indA{3}}\;{\scriptstyle (\mathbf1,\mathbf{35})}$};
\node (Y) at (4.5,-1.4) {$a^{(3)}\;{\scriptstyle (\mathbf1,\mathbf1)}$};
\foreach \u/\v in {B/S,C/P,R/S,R/P,R/T,H/T,S/A,P/A,T/A,P/D,T/D,T/Y}
  \draw[<->] (\u) to[bend left=7] (\v);
\end{tikzpicture} }
\caption{First copy.}
\label{fig:so37-boundary-first}
\end{subfigure}\hfill
\begin{subfigure}[t]{0.49\textwidth}
\centering
\resizebox{0.91\linewidth}{!}{%
\begin{tikzpicture}[x=1cm,y=1cm,
 every node/.style={draw,rounded corners=2pt,fill=white,inner xsep=5pt,inner ysep=3pt,minimum height=0.65cm},
 every path/.style={thick}]
\node (B) at (-4.5,1.4) {$b^{(6)}_{\indI{1}\indI{2}}\;{\scriptstyle (\mathbf5,\mathbf1)}$};
\node (C) at (-1.5,1.4) {$b^{(6)}_{\indA{1}\indA{2}}\;{\scriptstyle (\mathbf1,\mathbf{27})}$};
\node (R) at (1.5,1.4) {$b^{(6)}_{\indI{1}\indA{1}}\;{\scriptstyle (\mathbf3,\mathbf7)}$};
\node (H) at (4.5,1.4) {$b^{(6)}\;{\scriptstyle (\mathbf1,\mathbf1)}$};
\node (S) at (-3.0,0) {$\sigma_{-\,\indI{1}}\;{\scriptstyle (\mathbf4,\mathbf8)}$};
\node (P) at (0,0) {$\sigma_{-\,\indA{1}}\;{\scriptstyle (\mathbf2,\mathbf{48})}$};
\node (T) at (3.0,0) {$\theta_+\;{\scriptstyle (\mathbf2,\mathbf8)}$};
\node (A) at (1.5,-1.4) {$a^{(1)}_{\indI{1}\indA{1}\indA{2}}\;{\scriptstyle (\mathbf3,\mathbf{21})}$};
\node (D) at (-1.5,-1.4) {$a^{(1)}_{\indA{1}\indA{2}\indA{3}}\;{\scriptstyle (\mathbf1,\mathbf{35})}$};
\node (Y) at (4.5,-1.4) {$a^{(1)}\;{\scriptstyle (\mathbf1,\mathbf1)}$};
\foreach \u/\v in {B/S,C/P,R/S,R/P,R/T,H/T,S/A,P/A,T/A,P/D,T/D,T/Y}
  \draw[<->] (\u) to[bend left=7] (\v);
\end{tikzpicture} }
\caption{Second copy.}
\label{fig:so37-boundary-second}
\end{subfigure}
\caption{The two boundary coefficient multiplets for the polarized ${\rm SO}(3)\times {\rm SO}(7)$ background. Every line denotes a bidirectional supersymmetry connection.}
\label{fig:2multi37}
\end{figure}
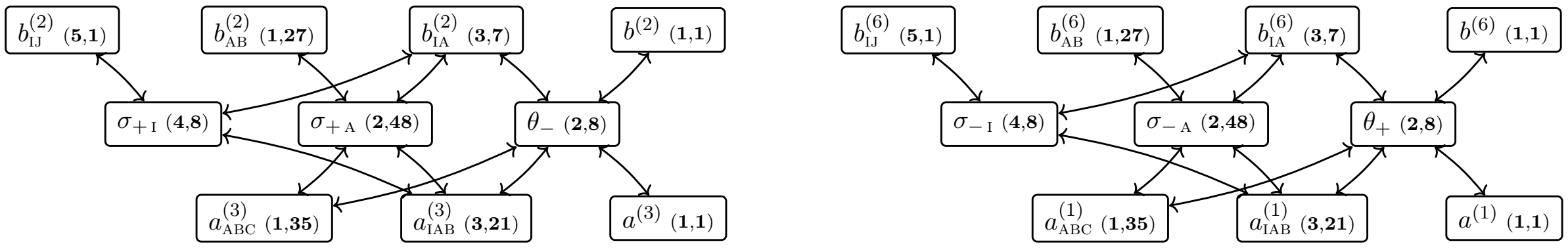

To summarize, we find that the bosonic and fermionic integration constants fall into two separate supermultiplets, just as in the ${\rm SO}(10)$ case. We sketch the resulting structure in Figure~\ref{fig:2multi37}. These multiplets no longer exhibit the nilpotent structure of the ${\rm SO}(10)$ multiplets, but rather reproduce the diamond structure of the $F(4)$ supermultiplet from Figure~\ref{fig:Bpolar}, describing the lowest operator content of the polarized IKKT model. Moreover, in contrast to the ${\rm SO}(10)$ case, here, the two multiplets are actually equivalent. 

Explicitly, after redefining 
\begin{align}
B_{\indI{1}\indI{2}}&=b^{(2)}_{\indI{1}\indI{2}}\,,\quad
 B_{\indA{1}\indA{2}}=b^{(2)}_{\indA{1}\indA{2}}\,,\quad
 B_{\indI{1}\indA{1}}=b^{(2)}_{\indI{1}\indA{1}}\,,
 \qquad
 \Psi_{\indI{1}}=\frac59\sigma_{+\,\indI{1}}\,,\;\;
 \Psi_{\indA{1}}=\frac59\sigma_{+\,\indA{1}}\,,\;\;
 \Psi=\frac59\theta_-\,,
 \nonumber\\
 A_{\indI{1}\indA{1}\indA{2}}&=
 \frac{\mathrm g}{4}a^{(3)}_{\indI{1}\indA{1}\indA{2}}\,,\quad
 A_{\indA{1}\indA{2}\indA{3}}=
 \frac{\mathrm g}{4}a^{(3)}_{\indA{1}\indA{2}\indA{3}}\,,\quad
 A=\frac{\mathrm g}{4}
 \bigl(a^{(3)}+\tfrac{32}{7}\hat\mu b^{(2)}\bigr)\,,\qquad
 S=
 \frac{\mathrm g}{4}
 \bigl(a^{(3)}+6\,\hat\mu b^{(2)}\bigr)
 \,,
  \label{eq:so37-rescale37-M1}
\end{align}
as in \eqref{eq:rescale-M1}, we recover precisely the $F(4)$ supermultiplet identified in the polarized IKKT model in section~\ref{subsec:F4}, with the identification
\begin{equation}
 \Omega=8i\mathrm g\,\hat\mu=i\frac{\mu}{2}
 \,,
  \label{eq:Omega-m}
\end{equation}
of the bulk and boundary deformation parameters.\footnote{The factor $i$ reflects that the real bulk axions $a_{ijk}$ are in fact dual to $i{\cal O}_{abc}$, whose bosonic part is real for Hermitian $X_a$, as in the coupling $i\Omega\,{\cal O}_\Omega$ of \eqref{eq:IKKTpolar}, while \eqref{eq:so37-rescale37-M1} and \eqref{eq:so37-rescale37-M2} pair them with ${\cal O}_{abc}$. Pairing them with $i{\cal O}_{abc}$ instead multiplies all ${\bf 120}$ data in \eqref{eq:so37-rescale37-M1}, including $S$, by $-i$, the ${\bf 144}$ data by $e^{-i\pi/4}$ and $\epsilon_0$ by $e^{i\pi/4}$, with inverse phases for the data in \eqref{eq:so37-rescale37-M2}. This replaces \eqref{eq:Omega-m} by the real identification $\Omega=8{\rm g}\hat\mu=\mu/2$.} Similarly, upon rescaling 
\begin{align}
\widetilde B_{\indI{1}\indI{2}}&=
 -6\mathrm g b^{(6)}_{\indI{1}\indI{2}},\;\;
 \widetilde B_{\indA{1}\indA{2}}=
 -6\mathrm g b^{(6)}_{\indA{1}\indA{2}},\;\;
 \widetilde B_{\indI{1}\indA{1}}=
 -6\mathrm g b^{(6)}_{\indI{1}\indA{1}},\quad
 \widetilde\Psi_{\indI{1}}=-\frac{27}{5}
 \sigma_{-\,\indI{1}}\,,\;\;
 \widetilde\Psi_{\indA{1}}=-\frac{27}{5}
 \sigma_{-\,\indA{1}}\,,\;\;
 \widetilde\Psi=-\frac{27}{5}\theta_+\,,
 \nonumber\\
&
 \widetilde A_{\indI{1}\indA{1}\indA{2}}=
 a^{(1)}_{\indI{1}\indA{1}\indA{2}}\,,\quad
 \widetilde A_{\indA{1}\indA{2}\indA{3}}=
 a^{(1)}_{\indA{1}\indA{2}\indA{3}}\,,\quad
\widetilde A =-\frac75\,a^{(1)}+\frac4{3\hat\mu}\, b^{(6)}\,,
\quad
  \widetilde S=
 -\frac{8}{\hat\mu}\left(b^{(6)}-\frac95\hat\mu a^{(1)}\right),
  \label{eq:so37-rescale37-M2}
\end{align}
as in \eqref{eq:rescale-M2}, we recover the dual multiplet, such that the pairing \eqref{eq:so10-dual-pairing} 
\begin{align}
 \mathcal I=\,&
  B_{\indI{1}\indI{2}}\widetilde B^{\indI{1}\indI{2}}
 +B_{\indA{1}\indA{2}}\widetilde B^{\indA{1}\indA{2}}
 +2B_{\indI{1}\indA{1}}\widetilde B^{\indI{1}\indA{1}}
  +\bar\Psi_{\indI{1}}\widetilde\Psi^{\indI{1}}
 +\bar\Psi_{\indA{1}}\widetilde\Psi^{\indA{1}}
 +\frac{10}{21}\bar\Psi\widetilde\Psi
 \nonumber\\
 &{}
 +3A_{\indI{1}\indA{1}\indA{2}}
       \widetilde A^{\indI{1}\indA{1}\indA{2}}
 +A_{\indA{1}\indA{2}\indA{3}}
       \widetilde A^{\indA{1}\indA{2}\indA{3}}
 +6A\widetilde A \;+\; S\tilde{S}
 \,,
 \label{eq:so37-final-dual-pairing}
\end{align}
is invariant under supersymmetry.  The last term combines the two supersymmetry invariants that decouple from the remaining multiplets.

After further rescaling, the multiplet \eqref{eq:so37-rescale37-M1} and its dual \eqref{eq:so37-rescale37-M2} are identical. The discussion of their ${\rm SO}(10)$ limit in the previous section suggests that also for the polarized model, the relevant sources should transform in the dual multiplet \eqref{eq:so37-rescale37-M2}.
In that case, this should also require a Legendre transform of the holographically renormalized on-shell action in the ${\bf 54}$ sector.


\section{Conclusions}
\label{sec:conclusions}

In this paper, we have presented the bulk realization of the lowest supermultiplet of BPS operators of the polarized IKKT matrix model. The polarization replaces the nilpotent supersymmetry structure of the undeformed IKKT model by the simple supergroup $F(4)$ with bosonic ${\rm SO}(3)\times{\rm SO}(7)$ symmetry. We have worked out how the lowest BPS multiplet of the undeformed model after proper deformation of the operators turns into the $L(2\delta)$ supermultiplet of $F(4)$. This requires the inclusion of Ward identities for the broken ${\rm SO}(10)$ symmetry generators in the ${\bf(3,7)}$, as well as the decoupling of a supersymmetry invariant linear combination of the scalar and axionic singlets. The resulting $L(2\delta)$ supermultiplet combines 152 bosonic and 144 fermionic states. After accounting for the global (super)symmetries, the resulting system describes 128 bosonic and 128 fermionic degrees of freedom.

On the bulk side, our main tool is the one-dimensional maximal supergravity obtained as a consistent truncation of Euclidean IIB supergravity on a nine-sphere $S^9$. Its fields precisely capture the operators of this multiplet and allow to study their proposed holographic realisation. We have analyzed its half-BPS domain walls and identified the unique ${\rm SO}(3)\times{\rm SO}(7)$-invariant branch whose leading deformation lies in the ${\bf 120}$ of ${\rm SO}(10)$, thus dual to the operator which defines the polarized IKKT model. This polarized background can be given in analytic form, including its Killing spinor, and provides the setting for the subsequent fluctuation analysis.

We have expanded the bulk field equations around both the ${\rm SO}(10)$-invariant and the polarized ${\rm SO}(3)\times{\rm SO}(7)$-invariant half-BPS domain walls. For the polarized background, after proper supersymmetry and gauge fixing, we have obtained analytic solutions for all physical linearized bosonic and fermionic fluctuations. The broken ${\rm SO}(10)$ transformations remove one mode in the mixed ${\bf (3,7)}$  sector, providing the bulk counterpart of the boundary Ward identities. Supersymmetry then organizes the independent asymptotic coefficients into two $F(4)$ multiplets. Their transformation rules reproduce the polarized IKKT multiplet and its dual upon identification of the bulk and boundary deformation parameters. The two boundary multiplets provide candidate operator and source data for polarized IKKT holography. In particular, the source assignment suggested by the undeformed limit identifies the ${\bf 54}$ scalar sources in the subleading branch, thus suggesting a further Legendre transform of the holographically renormalized on-shell action.

This sets the stage for systematically implementing the procedure of holographic renormalization following \cite{Bianchi:2001kw,Wiseman:2008qa,Kanitscheider:2008kd}, including suitable counterterms for the two asymptotic branches, thereby preparing the ground for the holographic computation of BPS correlators in the polarized IKKT model beyond one-point functions. The analytic fluctuation solutions can be followed throughout the bulk, possibly after adopting a more convenient global coordinate in place of $\tau$. Imposing appropriate conditions on them in the interior will, however, require an in-depth analysis of the regularity of the polarized background, whereas the present analysis only relied on its asymptotic UV behavior. In addition, with no boundary coordinate dependence to resolve, extracting correlation functions will hinge on carefully controlling the normalization of the bulk fields and their asymptotic modes. More generally, it would also be interesting to push this analysis to higher BPS multiplets, both on the matrix model and the bulk supergravity side.

Further aspects of IKKT holography have recently been explored through the matrix renormalization group flow \cite{Guan:2026sjw} and supersymmetric twists \cite{Hahner:2026mfg,Hahner:2026xpf}, and it would be worth relating these approaches to the present one. Another direction is to investigate the relation to the holographic D$(-1)$/D7 setup \cite{Billo:2021xzh}, whose near-horizon geometry shares features with some of the backgrounds found here \cite{Aguilar-Gutierrez:2022kvk}. Finally, it would be interesting to connect the holographic picture developed here to Monte Carlo studies of the dominant matrix configurations in the polarized model \cite{Chou:2025rwy} and, from a rather different perspective, to studies of emergent geometry in the IKKT model \cite{Steinacker:2016vgf,Ho:2025htr}.

\subsubsection*{Acknowledgements}

We thank Guillaume Bossard, Oleg Evnin, Juan Maldacena, and Xiang Zhao for useful discussions. We have used ChatGPT and Claude to improve the presentation of the manuscript and, together with Mathematica, for help with performing and checking some of the calculations. The authors take full responsibility for all results and conclusions.

\section*{Appendix}

\begin{appendix}

\section{Spinor conventions}
\label{app:spinor}

In this appendix, we collect our ${\rm SO}(10)$ spinor conventions. The $32\times 32$ complex gamma matrices $\Gamma_a$ satisfy the standard $\mathrm{SO}(10)$ Clifford algebra
\bea
\{\Gamma_a,\Gamma_b\}=2\,\delta_{ab}\,\mathbb{I}_{32}\,.
\eea
We choose a basis (c.f.\ the Appendix A of \cite{Komatsu:2024ydh}) in which 
$\Gamma_a^\dagger=\Gamma_a$ and the chirality gamma matrix $\Gamma_{*}$ is block diagonal,
\bea
\Gamma_{*}=-\mathrm{i}\,\Gamma_{1}\ldots\Gamma_{10}=\begin{pmatrix} \mathbb I_{16} &0 \\ 0 & -\mathbb I_{16}\label{eq:Gamma*} \end{pmatrix}\,.
\eea
It anticommutes with all gamma matrices and satisfies $\Gamma_{*}^2=\mathbb{I}_{32}$. Finally we choose a symmetric\footnote{This differs from \cite{Komatsu:2024ydh}, where the charge conjugation is antisymmetric and imaginary $\mathcal C_{\text{there}}=\mathrm{i}\,\mathcal C_{\text{here}}\,\Gamma_*$.} charge conjugation matrix,
\bea
\mathcal C=-\mathrm{i}\Gamma_{10}\Gamma_{*}=\begin{pmatrix}0 &\mathbb I_{16} \\ \mathbb I_{16} &0\end{pmatrix}\,,
\label{eq:CC}
\eea
which satisfies
\bea
\mathcal C=\bar{\mathcal C}=\mathcal C^T=\mathcal C^{-1}\,,\;\;\;\;\;\mathcal C^2=\mathbb{I}_{32}\,,\;\;\;\;\;\Gamma_a^T=\mathcal C\Gamma_a\mathcal C^{-1}\,.\label{eq:Csymm}
\eea
The last equation implies the following symmetry properties for antisymmetric products of gamma matrices contracted with $\mathcal C$,
\bea
\text{symmetric}&&:\;\mathcal C\,,\mathcal C\Gamma_a\,,\mathcal C\Gamma_a\Gamma_*\,,\mathcal C\Gamma_{ab}\Gamma_{*}\,,\mathcal C \Gamma_{abcd}\,,\mathcal C \Gamma_{abcde}\,,\nonumber\\
\text{antisymmetric}&&:\; \mathcal C\Gamma_{*}\,,\mathcal C\Gamma_{ab}\,,\mathcal C\Gamma_{abc}\,,\mathcal C\Gamma_{abc}\Gamma_{*}\,,\mathcal C\Gamma_{abcd}\Gamma_*\,.\label{eq:Gammasym}
\eea


 \section{Fluctuation equations}
\label{app:fluctuation}

In this appendix, we collect the explicit results for the dynamics of the fluctuations around the polarized ${\rm SO}(3)\times {\rm SO}(7)$ invariant background for the two sectors with non-trivial multiplicity: the ${\bf (1,1)}$ singlets, and the ${\bf (2,8)}$ fundamental spinors. In both cases, we give the explicit Lagrangians to quadratic order in the fluctuations.

\subsection{Singlet sector}
\label{subsec:singlet}
The Lagrangian for the four ${\bf (1,1)}$ singlet fluctuation modes $\{n, v, u, a\}$ 
around the polarized ${\rm SO}(3)\times{\rm SO}(7)$ background takes the form
\begin{align}
\mathrm g^{-1}\mathcal L^{(2)}_{t,\mathrm{sing}}={}&
80\bigl(\dot v^{2}-\tfrac1{21}\dot u^{2}\bigr)
-\frac{4\tau^{2/3}(1+\tau)^4(1+\tau^2)^{1/3}}
 {\hat\mu^{4/3}(1-\tau)^6}\dot a^{2}\nonumber\\
&+\frac{8\hat\mu^{8/3}(1-\tau)^2}
 {\tau^{4/3}(1+\tau)^2(1+\tau^2)^{2/3}}\,n
 \Bigl[\bigl(5(1+\tau^4)+12\tau(1+\tau^2)+20\tau^2\bigr)
 \dot v+3\tau(1+\tau^2)\dot u\Bigr]\nonumber\\
&+\frac{8\hat\mu^2(1-\tau+\tau^2)}
 {\sqrt{\tau(1+\tau^2)}}(n+2v-2u)\dot a
+\mathcal P_{uu}u^2+\mathcal P_{uv}uv
+\mathcal P_{vv}v^2+\mathcal P_{aa}a^2\nonumber\\
&+\mathcal P_{ua}ua+\mathcal P_{va}va
+\mathcal P_{un}un+\mathcal P_{vn}vn
+\mathcal P_{an}an+\mathcal P_{nn}n^2.
\label{eq:so37-singlet-component-action}
\end{align}
with the shorthand $\hat\mu=\mu/(16\mathrm{g})$. Here the coefficients $\mathcal P$ can be given as explicit functions of $\tau$. Explicitly,
\begin{align}
\mathcal P_{uu}={}&
\frac{\hat\mu^{16/3}(1-\tau)^4}
 {7\tau^{8/3}(1+\tau)^4(1+\tau^2)^{4/3}}
\bigl[5(1+\tau^8)-32\tau(1+\tau^6)
+308\tau^2(1+\tau^4)\nonumber\\
&\hspace{45mm}-304\tau^3(1+\tau^2)+718\tau^4\bigr]\,,
\nonumber\\
\mathcal P_{uv}={}&
\frac{16\hat\mu^{16/3}(1-\tau)^4}
 {\tau^{5/3}(1+\tau)^4(1+\tau^2)^{4/3}}
\bigl[7(1+\tau^6)+12\tau(1+\tau^4)
+29\tau^2(1+\tau^2)+12\tau^3\bigr]\,,
\nonumber\\
\mathcal P_{vv}={}&
\frac{16\hat\mu^{16/3}(1-\tau)^4}
 {\tau^{8/3}(1+\tau)^4(1+\tau^2)^{4/3}}
\bigl[10(1+\tau^8)+29\tau(1+\tau^6)
+54\tau^2(1+\tau^4)\nonumber\\
&\hspace{45mm}+88\tau^3(1+\tau^2)+124\tau^4\bigr]\,,
\nonumber\\
\mathcal P_{aa}={}&
\frac{3\hat\mu^4(1-\tau)^4(1+\tau)^2}
 {16\tau^2(1+\tau^2)}\,,
\nonumber\\
\mathcal P_{ua}={}&
\frac{2\hat\mu^{14/3}(1-\tau)^4(1+\tau+\tau^2)}
 {\tau^{11/6}(1+\tau^2)^{7/6}}\,,\nonumber\\
\mathcal P_{va}={}&
\frac{18\hat\mu^{14/3}(1-\tau)^4(1+\tau+\tau^2)}
 {\tau^{11/6}(1+\tau^2)^{7/6}}\,,
\nonumber\\
\mathcal P_{un}={}&
\frac{2\hat\mu^{16/3}(1-\tau)^4}
 {\tau^{5/3}(1+\tau)^4(1+\tau^2)^{4/3}}
\bigl[11(1+\tau^6)+4\tau(1+\tau^4)
+61\tau^2(1+\tau^2)-8\tau^3\bigr]\,,
\nonumber\\
\mathcal P_{vn}={}&
\frac{8\hat\mu^{16/3}(1-\tau)^4}
 {\tau^{8/3}(1+\tau)^4(1+\tau^2)^{4/3}}
\bigl[5(1+\tau^8)+16\tau(1+\tau^6)
+39\tau^2(1+\tau^4)\nonumber\\
&\hspace{45mm}+56\tau^3(1+\tau^2)+92\tau^4\bigr]\,,
\nonumber\\
\mathcal P_{an}={}&
\frac{3\hat\mu^{14/3}(1-\tau)^4(1+\tau+\tau^2)}
 {\tau^{11/6}(1+\tau^2)^{7/6}}\,,\nonumber\\
\mathcal P_{nn}={}&
\frac{\hat\mu^{16/3}(1-\tau)^4(5+17\tau^2+5\tau^4)}
 {\tau^{8/3}(1+\tau^2)^{4/3}}\,.
\label{eq:so37-singlet-potential-coefficients}
\end{align}

\subsection{Fundamental spinor sector}
\label{subsec:spinor}
The Lagrangian for the three fermionic fluctuation modes $\{\psi, \lambda, \Theta\}$ in the fundamental spinor representation ${\bf (2,8)}$ 
around the polarized ${\rm SO}(3)\times{\rm SO}(7)$ background takes the form
\begin{align}
\mathcal L^{(2)}_{(\mathbf2,\mathbf8)}={}&
20\bar\lambda\dot\lambda+\frac{20}{21}\bar\Theta\dot\Theta
\label{eq:so37-28-lagrangian}
\\
&
+\frac{\hat\mu^{8/3}(1-\tau)^2}
 {8\tau^{4/3}(1+\tau)^2(1+\tau^2)^{2/3}}
\Bigl[
\bar\lambda\mathsf M_{\lambda\lambda}\lambda
+\bar\lambda\mathsf M_{\lambda\Theta}\Theta
+\bar\Theta\mathsf M_{\Theta\Theta}\Theta
+64\mathrm g\,\bar\psi
\bigl(\mathsf M_{\psi\lambda}\lambda+
      \mathsf M_{\psi\Theta}\Theta\bigr)
\Bigr]\,,
\nonumber
\end{align}
where all five mass terms can be given in explicit polynomials in $\tau$ as
\begin{align}
\mathsf M_{\lambda\lambda}={}&
-42(1-\tau^2)(5+4\tau+5\tau^2)\Gamma_*
-48\tau^{1/2}(1-\tau)(1-\tau+\tau^2)\Gamma_{\#}\nonumber\\
&-336\tau^{1/2}(1+\tau)(1+\tau+\tau^2)
 \Gamma_{\#}\Gamma_*\,,
\nonumber\\[1ex]
\mathsf M_{\lambda\Theta}={}&
-64\tau(1-\tau^2)
-48\tau^{1/2}(1+\tau)(1+\tau+\tau^2)\Gamma_{\#}
+16\tau^{1/2}(1-\tau)(1-\tau+\tau^2)\Gamma_{\#}\Gamma_*\,,
\nonumber\\[1ex]
\mathsf M_{\Theta\Theta}={}&
-\frac{2}{21}(1-\tau^2)(5+36\tau+5\tau^2)\Gamma_*
-\frac{128}{21}\tau^{1/2}(1-\tau)(1-\tau+\tau^2)\Gamma_{\#}
\nonumber\\
&
+\frac{64}{21}\tau^{1/2}(1+\tau)(1+\tau+\tau^2)
 \Gamma_{\#}\Gamma_*\,,
\nonumber\\[1ex]
\mathsf M_{\psi\lambda}={}&
(1-\tau^2)(5+4\tau+5\tau^2)
+(5\tau^4+12\tau^3+20\tau^2+12\tau+5)\,\Gamma_*
\nonumber\\
&
-9\tau^{1/2}(1+\tau)(1+\tau+\tau^2)\Gamma_{\#}
-\tau^{1/2}(1-\tau)(1-\tau+\tau^2)
 \Gamma_{\#}\Gamma_*\,,
\nonumber\\[1ex]
\mathsf M_{\psi\Theta}={}&
3\tau(1+\tau^2)+\tau(1-\tau^2)\Gamma_*
+\tau^{1/2}(1-\tau)(1-\tau+\tau^2)\Gamma_{\#}
\nonumber\\
&
-\tau^{1/2}(1+\tau)(1+\tau+\tau^2)
 \Gamma_{\#}\Gamma_*\,.
\label{eq:so37-28-lagrangian-matrices}
\end{align}
We note that the gravitino $\psi$ enters
the Lagrangian only through the negative-chirality combination
$\psi_-+\tau^{1/2}\Gamma_{\#}\psi_+$, \textit{i.e.}\ its field equation has only sixteen
independent components. This
is the same rank as in the ${\rm SO}(10)$ background, where $\psi_+$
was absent from the Lagrangian altogether.
Equivalently, the two matrices $ \mathsf M_{\psi\lambda}$, $\mathsf M_{\psi\Theta}$,
obey the left-projection constraint
\begin{equation}
 \bigl(\mathbb{P}_-+\tau^{1/2}\Gamma_{\#}\mathbb{P}_+\bigr)
 \mathsf M_{\psi\lambda}=0\,,
 \qquad
 \bigl(\mathbb{P}_-+\tau^{1/2}\Gamma_{\#}\mathbb{P}_+\bigr)
 \mathsf M_{\psi\Theta}=0\,,
 \label{eq:so37-28-matrix-projection}
\end{equation}
with the chiral projectors
\begin{equation}
\mathbb{P}_\pm=\frac12\,(\mathbb{I} \pm \Gamma_*)\,.
\end{equation}
This can be checked directly against
\eqref{eq:so37-28-lagrangian-matrices}: the two matrices factorize as
\begin{align}
\mathsf M_{\psi\lambda}={}&
 (1-\tau^{1/2}\Gamma_{\#})
 \bigl[2(5+8\tau+10\tau^2+4\tau^3)\mathbb{P}_+
 -2\tau^{1/2}(4+10\tau+8\tau^2+5\tau^3)
 \Gamma_{\#}\mathbb{P}_-\bigr]\,,
 \nonumber\\
\mathsf M_{\psi\Theta}={}&
 2\,(1-\tau^{1/2}\Gamma_{\#})
 \bigl[\tau(2+\tau^2)\mathbb{P}_+
 +\tau^{1/2}(1+2\tau^2)\Gamma_{\#}\mathbb{P}_-\bigr]\,.
 \label{eq:so37-28-matrix-factorization}
\end{align}

\end{appendix}


\providecommand{\href}[2]{#2}\begingroup\raggedright\endgroup

\end{document}